\documentclass[%
 reprint,
 amsmath,amssymb,
 aps,
]{revtex4-2}

\usepackage{graphicx}
\usepackage{bm,booktabs}

\begin{document}

\preprint{APS/123-QED}

\title{Black Holes as Non-Abelian Anyon Condensates:\\
Implications for the Information Paradox}

\author{Sabin Roman}
\affiliation{%
Department of Knowledge Technologies, Jo\v{z}ef Stefan Institute, Slovenia
}%


\begin{abstract}
We propose a black-hole model in which the would-be horizon is replaced by a topologically ordered timelike shell of condensed non-Abelian anyons surrounding a regular, topologically trivial vacuum interior. Effective bosonization of the underlying matter produces a $(2+1)$-dimensional phase in which area is the natural extensive variable. The shell microstates form a constrained fusion Hilbert space reproducing the Bekenstein--Hawking entropy, while discrete area and mass spectra generate logarithmic and inverse-area thermodynamic corrections. Equipartition recovers the Hawking temperature, and a quadratic collective Hamiltonian reproduces the logarithmic canonical entropy and connects collective modes to candidate anyon quantum dimensions. We use Einstein--Weyl gravity as an effective high-curvature description of the transition region, in which a Planck-thick membrane consistently joins a regular constant-curvature interior to a Schwarzschild exterior near the would-be horizon. In contrast to the corresponding Israel shell in general relativity, whose evolution ends in collapse, the Einstein--Weyl shell possesses a linearly stable radial regime for both anisotropic and isotropic response, whose upper stability boundary coincides with the Buchdahl limit. The discrete anyon spectrum determines transition weights which, under a fluctuation--dissipation mapping, imply absorptivity and an effective shear response approaching the classical horizon limit at high frequency. Inward Hawking radiation is reabsorbed rather than accumulated into a Tolman equilibrium bath, while reflectivity can generate gravitational-wave echoes. Information is encoded nonlocally in fusion channels and may be released through shell-state transitions without trans-horizon entanglement, casting the black hole as a quantum-information system grounded in topological quantum computation.
\end{abstract}

\maketitle

\section{Introduction}

The standard classical picture of black-hole formation assumes that once all
forms of degeneracy pressure are overcome, gravitational collapse proceeds to a
trapped region and, in classical general relativity under standard assumptions,
to a singularity hidden behind an event horizon
\cite{Penrose1965,Wald1984}. We propose an alternative scenario: once collapse
passes beyond the neutron-degeneracy limit, matter undergoes a phase transition
into a two-dimensional condensate of non-Abelian anyons rather than continuing
to a singularity. These quasiparticles arise in topological phases of matter
\cite{wilczek1982,kitaev2003fault} and, in the present model, provide a
natural route to an effective bosonization of underlying fermionic degrees of
freedom, distinct from conventional Bose--Einstein condensation. Their non-Abelian fusion rules
generate a constrained many-body Hilbert space and allow quantum information to
be stored nonlocally in fusion channels
\cite{levin2005string,nayak2008non}. In the present picture, collapse therefore
terminates not in a pointlike singular state but in a topologically ordered
many-body phase localized on a timelike shell.

Such a transition is expected only in an extreme, Planckian regime, where
density and pressure lie beyond the reach of standard effective descriptions.
In the same regime, it is natural to expect the Einstein--Hilbert description
of gravity to receive ultraviolet corrections. More generally, once curvature
or local variation scales approach the Planck scale, higher-curvature terms
should enter the effective action
\cite{donoghue1994general,donoghue2012effective}. From this perspective,
combinations of \(R\), \(R^2\), and
\(C_{\mu\nu\rho\sigma}C^{\mu\nu\rho\sigma}\) provide a natural effective
setting for high-curvature gravitational dynamics
\cite{stelle1977renormalization}. This motivates the Einstein--Weyl theory as a minimal local extension of
general relativity capable of supporting the higher-derivative response
required in the transition region. The Weyl-squared term becomes increasingly
relevant at short transverse scales, but the membrane relations used below are
derived from the full Einstein--Weyl equations. The higher-derivative sector is
used here as an effective classical tool rather than as a fundamental quantum
theory, and ghost-related issues of a fundamental Weyl-squared theory are not
addressed.

The resulting shell need not disappear once the exterior geometry returns to
the ordinary Einstein regime. The distinction is one of scale. For a
macroscopic black hole the exterior varies on scales of order
\(M\gg\ell_P\), so higher-derivative corrections are strongly suppressed there.
Across the condensate, however, the proper transverse thickness may remain
\(h=O(\ell_P)\). Schematically, the Einstein and Weyl contributions scale as
\(M_P^2/L^2\) and \(\alpha_g/L^4\), respectively, so Planck-scale transverse
gradients can keep the higher-derivative response relevant inside the membrane
even while the surrounding bulk is accurately described by general
relativity. The shell may therefore be viewed as a localized ultraviolet
remnant of the phase reached during collapse, embedded in a spacetime that is well described by Einstein gravity at larger scales.

A familiar condensed-matter example is an Abrikosov vortex in a type-II
superconductor. Its core is microscopic and depends on the short-distance
condensate physics, whereas on larger scales the vortex persists as an
effective classical defect carrying quantized magnetic flux and nontrivial
winding \cite{Abrikosov1957,Mermin1979}. Related, although more hypothetical,
examples occur in relativistic field theory, where vortex strings, cosmic
strings, and magnetic monopoles produced during high-energy phase transitions
can survive as localized defects in the subsequent low-energy theory
\cite{NielsenOlesen1973,Kibble1976,tHooft1974,Polyakov1974}. The analogy is not
that the anyonic shell is itself such a defect, but that microscopic
high-energy physics can leave a localized remnant whose collective dynamics
survive in a different long-wavelength regime. In the present case, topological order provides an additional source of
robustness: the internal fusion information is protected against local
perturbations \cite{kitaev2003fault,nayak2008non}, although the lifetime and
metastability of the condensate must ultimately follow from its microscopic
dynamics.

A further point of contact between the anyonic description and conformal
gravity is structural. In nonlinear geometric theories, a pointlike source is
considerably less natural than in a linear field theory: superposition is lost,
and singular delta-function sources need not possess the simple Green-function
interpretation familiar from electrodynamics. Higher-derivative gravitational equations are instead sensitive to the spatial
structure and distributional moments of localized sources, including
double-layer contributions beyond the standard thin-shell structure of
Einstein gravity \cite{ReinaSenovillaVera2016,
BerezinDokuchaevEroshenko2019}. Thin shells and finite surface layers therefore
provide a more controlled setting for localizing stress--energy on
codimension-one hypersurfaces. In this spirit, the present work models the
black-hole degrees of freedom as an extended shell rather than as an effective
pointlike interior object.

In this picture, the anyonic condensate nucleates at the core and grows outward
during collapse until it forms a timelike shell near the would-be horizon,
whose area is postulated to match that of the corresponding black hole. Its
non-Abelian fusion degrees of freedom provide a finite constrained Hilbert
space in which quantum information is stored nonlocally
\cite{kitaev2003fault,nayak2008non}. Because non-Abelian anyons are
intrinsically associated with a two-dimensional spatial manifold, the
\((2+1)\)-dimensional intrinsic geometry of a timelike shell provides the
natural arena for their braiding and fusion structure. The would-be horizon
is therefore treated not merely as a causal boundary, but as a physical
substrate whose number of microscopic constituents is extensive in area.
The Bekenstein--Hawking entropy consequently acquires a direct microscopic
interpretation as the entropy of a physical two-dimensional system, while
transitions between its fusion states provide a natural channel for
information-carrying radiation. This retains the area-based intuition
underlying holographic descriptions
\cite{hooft1993dimensional,susskind1995world}, but does not require
bulk--boundary duality or the holographic principle as an independent
microscopic postulate.

A central difficulty with taking such an entropy-bearing shell literally in
general relativity is its dynamics. Self-gravitating material shells are
strongly constrained, and recent perturbative work finds nonradial
instabilities for the direct Minkowski-interior/Schwarzschild-exterior
benchmark \cite{PitreSchneiderPoisson2026}. The radial Israel problem is also
increasingly restrictive as the shell approaches the Schwarzschild radius.
The Einstein--Weyl membrane changes this behavior: the higher-derivative
response allows a finite positive-tension restoring channel in the
near-horizon regime. The allowed radial regime extends through the
ultracompact region and ends at the familiar Buchdahl compactness scale. This
is physically consistent with the intended interpretation of the object: it
is a black-hole replacement, not an ordinary compact star, and is therefore
expected to reside on the more compact side of the stellar Buchdahl radius.
The result remains a conditional statement about the leading radial mode;
nonradial stability is a separate question.

Because the shell is timelike, mechanical support alone does not determine
whether it behaves observationally like a horizon. The same fusion degeneracy
and discrete mass spectrum used for the thermodynamics also fix a
discrete sequence of transition frequencies and reverse/absorption rate
ratios. When these microscopic results are matched to the generic
fluctuation--dissipation response of a quantum horizon, they imply a
frequency-dependent line reflectivity and an effective membrane-equivalent shear
response, while strongly overlapping lines recover the smooth Boltzmann
envelope \cite{thorne1986black,OshitaWangAfshordi2020}. This also gives the
regular interior a consistent nonequilibrium interpretation: Hawking-order
quanta emitted inward are reabsorbed after only a few crossings rather than
building an equilibrium Tolman bath, while residual line reflectivity can
source gravitational-wave echoes.

The same microscopic picture changes the formulation of the black-hole
information problem. In the semiclassical description, Hawking radiation
\cite{hawking1975particle} leads to a tension between approximately thermal
emission and quantum unitarity, motivating proposals involving trans-horizon
entanglement, complementarity and firewalls \cite{almheiri2013black}, as well
as later Page-curve and replica-wormhole constructions
\cite{AlmheiriEtAl2020Replica}. Here the microscopic Hilbert space is instead
localized on the timelike condensate from the outset, and radiation is
described by thermally weighted transitions between discrete shell states,
whose amplitudes and selection rules may depend on the fusion-channel state.
There is therefore no inaccessible interior partner Hilbert space from which
information must later be reconstructed, and the standard information paradox
does not arise in its usual form. Information remains associated with the
physical shell degrees of freedom throughout the evolution and can, in
principle, be carried into the radiation through the sequence of shell-state
transitions. The remaining issue is dynamical rather than paradoxical: whether
the complete microscopic transition amplitudes of the condensate define a
unitary, information-preserving evolution and ultimately transfer the encoded
fusion information consistently to the outgoing radiation.

Section~\ref{sec:model} develops the condensate thermodynamics. Fusion-state
counting gives the leading area entropy and fixes the discrete area and mass
spectra, including logarithmic and inverse-area corrections. The resulting
scale separation then justifies an equipartition treatment of the local shell
modes, which recovers the black-hole temperature for Schwarzschild, Kerr,
Reissner--Nordstr\"om, and Kerr--Newman geometries. Building on that thermal
scale, a minimal collective Hamiltonian is introduced and its canonical entropy
is shown to reproduce the leading and logarithmic structure. Section~\ref{sec:formation}
develops the gravitational realization. A regular constant-curvature interior
is connected through a finite-width anisotropic membrane to a Schwarzschild-
like exterior; the Einstein--Weyl equations determine the leading source
moment and the effective radial dynamics give a conditional near-horizon
stability criterion. The same microscopic spectrum is then used to derive the
discrete transition weights and line spacing, infer the corresponding
coarse-grained absorptive response and membrane-equivalent shear response,
test the regular interior against Tolman accumulation, and determine the
leading echo delay and suppression. Section~\ref{sec:discussion}
discusses the information problem, relations to other microscopic black-hole
proposals, and observational implications. Appendix~\ref{app:entropy} collects
notation and the detailed canonical-entropy calculation, while
Appendix~\ref{app:radial_stability} gives the full Israel and Einstein--Weyl
radial-stability derivations.

\section{Thermodynamics and Effective Hamiltonian}
\label{sec:model}

In this section we develop the thermodynamic and microscopic consequences of
modeling the horizon as a condensate of non-Abelian anyons localized on a thin
shell. We proceed in three steps. First, fusion-state counting together with a
discrete area spectrum fixes the entropy, mass levels, transition scale, and
subleading thermodynamic corrections. Second, the resulting hierarchy between
the global level spacing and the local excitation scale justifies a classical
equipartition treatment of the coarse-grained shell modes, which reproduces the
Hawking temperature. Third, using that thermal behavior as motivation, we
introduce a minimal collective Hamiltonian and test whether its canonical
entropy reproduces the same leading and logarithmic structure.

\subsection{Fusion entropy and black hole thermodynamics}

Within the framework we introduce, the entropy of the black hole arises from
the dimension of the constrained fusion Hilbert space of the anyons localized
on the shell. Assuming a trivial total topological charge and a large anyon
number \(N\), the fusion Hilbert space grows asymptotically as
\(\dim\mathcal H_N\propto d^N\), where \(d\) is the quantum dimension of the
anyon species \cite{nayak2008non}. The corresponding fusion entropy is therefore
\begin{equation}
    \boxed{
    S_{\rm fus}(N)
    =
    N\log d + O(1).
    }
\label{eq:Sfus}
\end{equation}
The leading term counts the exponential growth of fusion-consistent states,
which serve as the microstates underlying the macroscopic entropy. Fusion
constraints, topology, and boundary conditions contribute only subleading terms
and therefore do not modify the extensive large-\(N\) scaling.

Since the microscopic degrees of freedom are confined to a two-dimensional
shell, an area law for the entropy is the natural extensive scaling. We identify
\(A=NA_0\), where \(A_0\) is the effective area associated with one
constituent. The macroscopic area is therefore built from a discrete surface
Hilbert space, providing a microscopic realization of the type of area
quantization proposed in early work~\cite{Bekenstein1974KerrSpectrum}.

Using \(N=A/A_0\), Eq.~\eqref{eq:Sfus} becomes
\(S_{\rm fus}(A)=\alpha A+O(1)\), with
\(\alpha=\log d/A_0\). Matching the leading term to the
Bekenstein--Hawking entropy,
\(S_{\rm BH}=A/4\)
\cite{bekenstein1973black,hawking1975particle}, implies
\(\alpha=1/4\) and fixes
\begin{equation}
    \boxed{
    A_0
    =
    4\log d.
    }
\label{eq:A0}
\end{equation}
Thus the quantum dimension directly determines the microscopic area scale
associated with each constituent. Since the ADM mass inferred from the asymptotic Schwarzschild exterior obeys
\(A=16\pi M^2\) in Planck units, the discrete area spectrum motivates
\begin{equation}
    M_n
    =
    \mu\sqrt n,
    \qquad
    n\in\mathbb Z_{\geq1}.
\end{equation}
Here \(n\) labels successive area levels. The mass scale is then
\(\mu^2=A_0/(16\pi)=\log d/(4\pi)\). The resulting mass spacing
is therefore fixed by the same quantum dimension that controls the asymptotic
growth of the fusion Hilbert space. The energy of an emitted quantum is
\begin{equation}
    \Delta E_n
    =
    M_n-M_{n-1}
    =
    \mu\left(\sqrt n-\sqrt{n-1}\right).
\label{eq:deltaE}
\end{equation}
Using \(\Delta S=\log d\), the effective temperature associated with a single
transition is
\begin{align}
    T_n
    &=
    \frac{\Delta E_n}{\Delta S}
    =
    \frac{\mu}{\log d}
    \left(\sqrt n-\sqrt{n-1}\right)
    \notag\\
    &=
    \frac{\mu}{\log d}
    \frac{1}{2\sqrt n}
    \left(
        1+\frac{1}{4n}+\frac{1}{8n^2}+\cdots
    \right).
\label{eq:T}
\end{align}

The elementary transition also admits a direct information-theoretic interpretation. Since \(g_n/g_{n-1}\simeq d\), a transition \(n\rightarrow n-1\) reduces the fusion-space entropy by \(\Delta S=\log d\). Landauer's principle associates the quasistatic erasure of this amount of information at temperature \(T\) with the minimum energy cost \(T\Delta S\)~\cite{Landauer1961,Bennett2003}. Equation~\eqref{eq:T} therefore gives \(\Delta E_n=T_n\log d\), so that an elementary shell transition has precisely the Landauer form. Equivalently, the area step \(A_0=4\ell_P^2\log d\) generalizes the familiar binary \(4\ell_P^2\log 2\) spacing to a fusion space of quantum dimension \(d\).

This yields the Hawking temperature with controlled corrections:
\begin{equation}
    \boxed{
    T(M)
    =
    \frac{1}{8\pi M}
    \left[
        1
        +
        \frac{\log d}{16\pi M^2}
        +
        \frac{(\log d)^2}{128\pi^2M^4}
        +
        \cdots
    \right]
    }.
\end{equation}
which reduces to the standard result in the macroscopic limit.

Integrating \(dS/dM=1/T(M)\) recovers the entropy:
\begin{align}
    S_{\mathrm{therm}}(M)
    &=
    4\pi M^2
    -
    \frac{\log d}{2}\log M
    +
    \frac{(\log d)^2}{64\pi M^2}
    +
    \cdots ,
    \\
    &\boxed{
    \begin{aligned}
    S_{\mathrm{therm}}(A)
    ={}&
    \frac{A}{4}
    -
    \frac{\log d}{4}\log A
    \\
    &+
    \frac{(\log d)^2}{4A}
    +
    \cdots
    \end{aligned}
    }.
\label{eq:Stherm}
\end{align}

This thermodynamic entropy \(S_{\mathrm{therm}}\) reproduces the leading term
obtained from microstate counting and admits logarithmic and inverse-area
corrections. Similar structures arise in treatments based on quantized
black-hole spectra and simple degeneracy assumptions. Bekenstein and
Mukhanov~\cite{BekensteinMukhanov1995Spectroscopy} considered a uniformly
spaced area spectrum with fixed degeneracy per quantum, while later work
allowed the spacing and degeneracy to vary independently
\cite{Oppenheim2003SpectrumQNM}. In both cases, the Bekenstein--Hawking law
constrains the relation between the spectrum and its degeneracy.

Related calculations based on discrete area or mass levels also produce
logarithmic corrections \cite{Blaschke2017,HeCai2021}, with logarithmic and
inverse-area terms obtained explicitly in Ref.~\cite{JanaDuttaGangopadhyay2026}. These results are
obtained through somewhat different discrete or continuum treatments, but
their common structure supports the interpretation of such corrections as a
generic consequence of quantized black-hole thermodynamics rather than of one
particular calculational prescription.

Logarithmic corrections also arise from statistical and ensemble
fluctuations
\cite{DasMajumdarBhaduri2002GeneralLog,BhaduriTranDas2004Microcanonical},
and in approaches based on loop quantum gravity
\cite{ashtekar1998quantum,kaul2000logarithmic}
and Euclidean quantum-gravity calculations~\cite{sen2013logarithmic}. The present calculation should
therefore not be viewed as deriving a unique correction, but as reproducing a
well-established hierarchy within a specific microscopic shell model.

What is more distinctive here is that the same hierarchy is naturally
compatible with non-Abelian fusion data. The quantum dimension \(d\) controls
the exponential growth of the fusion Hilbert space, while global fusion
constraints provide subleading structure
\cite{nayak2008non}. Related anyonic and Chern--Simons descriptions of
horizon states have appeared in loop quantum gravity
\cite{ashtekar2000isolated,pithis2015anyonic}, but the connection between
fusion-space growth, a discrete black-hole spectrum, and the resulting
thermodynamic corrections has not usually been developed as a single
mechanism. In the present model, these elements fit together without requiring
separate microscopic assumptions for the leading entropy and its first
corrections.

\subsection{Equipartition and the Hawking temperature}
\label{sec:thermodynamics}

Although the global spectrum is discrete, the local shell modes need not be quantum mechanically sparse. The anyonic shell is timelike and lies just outside the would-be Schwarzschild horizon, \(r_c>r_s=2M\), so it is not separated from the exterior by an event horizon: outward-directed radiation can propagate directly from the shell to infinity. Radiation may therefore be described as ordinary thermodynamic emission through discrete, thermally weighted transitions between shell microstates, without invoking quantum tunnelling or field-theoretic pair creation across a horizon. For a thermally active transition, the characteristic level spacing is generically of order the thermal scale, $\Delta E_n=O(T_n)$. If the energy of such a global transition is distributed over the $N$ shell constituents, the corresponding scale per constituent is \begin{equation} \boxed{ \frac{\Delta E_n}{N} = O\!\left(\frac{T_n}{N}\right) = O\!\left(\frac{T_n}{M^2}\right) \ll T_n } \label{eq:equipartition_scale_separation} \end{equation} for $M\gg1$, since $N\propto A\propto M^2$. Thus the black-hole spectrum remains discrete at the global level, while each local or coarse-grained degree of freedom lies deep in the thermally populated regime. Quantum effects enter through the global
constraints and discrete transitions, but the collective thermodynamics of
the shell is effectively classical.

The coexistence of a discrete global spectrum with effectively classical
collective thermodynamics is familiar in quantum many-body systems. Lattice
vibrations remain quantized as phonons, but highly occupied low-frequency
modes obey classical equipartition. Likewise, coarse-grained magnetization
modes and collective modes of quantum condensates can be treated classically
when their local excitation spacing is small compared with the thermal scale.
The anyonic shell is intended in the same sense: its fusion constraints and
global transitions remain quantum, while its local collective modes admit a
classical equipartition description.

The preceding scale separation shows that the local excitation scale is small compared with the temperature even though the global transition spectrum remains discrete. We assign one effective degree of freedom per Planck area $\ell_P^2$. The total number of such degrees of freedom is then
\(N_{\rm dof}:=A/\ell_P^2\). For a Schwarzschild black hole, the effective
thermal energy is the total mass, \(E_{\rm therm}=M\). Applying
\(E_{\rm therm}=\tfrac12N_{\rm dof}T\) gives
\begin{equation}
    \boxed{
    T
    =
    \frac{2E_{\rm therm}}{N_{\rm dof}}
    =
    \frac{2M}{A}
    =
    \frac{1}{8\pi M},
    }
\label{eq:temp}
\end{equation}
which recovers the Hawking temperature in an elementary way.

For rotating and charged configurations, write the thermal energy as
\(E_{\rm therm}=M f(x)\), where \(x\) is the irreducible mass normalized by
its extremal value within the corresponding one-parameter family.

For a Kerr black hole, define \(a:=J/M^2\). The irreducible mass satisfies
\[
    M_{\rm irr}^2
    =
    \frac{A_{\rm Kerr}}{16\pi}
    =
    \frac{M^2}{2}
    \left(
        1+\sqrt{1-a^2}
    \right).
\]
Since the extremal Kerr value is
\(M_{\rm irr}^{\rm ext}=M/\sqrt{2}\), introduce
\(x_{\rm Kerr}:=M_{\rm irr}/(M/\sqrt{2})\). The exact irreducible-mass
relation is \(x_{\rm Kerr}^2=1+\sqrt{1-a^2}\).

The thermal-energy fraction must vanish at extremality and equal unity in the
Schwarzschild limit. These limits correspond to \(x_{\rm Kerr}=1\) and
\(x_{\rm Kerr}=\sqrt{2}\), respectively, so the endpoint conditions are
\(f_{\rm Kerr}(1)=0\) and \(f_{\rm Kerr}(\sqrt{2})=1\). Since the exact
irreducible-mass relation is linear in \(x_{\rm Kerr}^2\), the minimal affine
form \(f_{\rm Kerr}(x)=c_1x^2+c_0\) is appropriate. The endpoint conditions
fix \(c_1=1\) and \(c_0=-1\), and hence
\(f_{\rm Kerr}(x)=x^2-1\). It follows that
\begin{equation}
    \boxed{
    E_{\rm therm}^{\rm Kerr}
    =
    M\left(x_{\rm Kerr}^2-1\right)
    =
    M\sqrt{1-a^2}.
    }
\label{eq:kerr_thermal_energy}
\end{equation}
Together with
\(A_{\rm Kerr}=8\pi M^2\left(1+\sqrt{1-a^2}\right)\), the equipartition
relation reproduces the Kerr Hawking temperature.

For a Reissner--Nordstr\"om black hole, define \(q:=Q/M\). Its irreducible
mass is
\[
    M_{\rm irr}
    =
    \frac{r_+}{2}
    =
    \frac{M}{2}
    \left(
        1+\sqrt{1-q^2}
    \right).
\]
Since the extremal value is \(M_{\rm irr}^{\rm ext}=M/2\), introduce
\(x_{\rm RN}:=M_{\rm irr}/(M/2)\). The exact irreducible-mass relation is
\(x_{\rm RN}=1+\sqrt{1-q^2}\).

Here extremality corresponds to \(x_{\rm RN}=1\), while the Schwarzschild
limit corresponds to \(x_{\rm RN}=2\). The endpoint conditions are therefore
\(f_{\rm RN}(1)=0\) and \(f_{\rm RN}(2)=1\). Since the irreducible-mass
relation is linear in \(x_{\rm RN}\), take the minimal affine form
\(f_{\rm RN}(x)=c_1x+c_0\). The endpoint conditions again give \(c_1=1\)
and \(c_0=-1\), so that \(f_{\rm RN}(x)=x-1\). It follows that
\begin{equation}
    \boxed{
    E_{\rm therm}^{\rm RN}
    =
    M\left(x_{\rm RN}-1\right)
    =
    M\sqrt{1-q^2}.
    }
\label{eq:rn_thermal_energy}
\end{equation}
Together with
\(A_{\rm RN}=4\pi M^2\left(1+\sqrt{1-q^2}\right)^2\), equipartition again
reproduces the corresponding Hawking temperature.

The quadratic Kerr relation and linear Reissner--Nordstr\"om relation thus
follow from their respective irreducible-mass dependences and endpoint
conditions. In each one-parameter family, the construction interpolates
between \(E_{\rm therm}=M\) in the Schwarzschild limit and
\(E_{\rm therm}=0\) at extremality.

For a generic Kerr--Newman black hole, however, the irreducible mass alone
does not provide an equally simple one-variable construction. The extremal
value of \(M_{\rm irr}/M\) depends on the relative contributions of spin and
charge, so there is no unique normalized variable \(x\) for which a single
polynomial \(f(x)\) extends both of the preceding arguments.

The effective thermal energy that yields the Kerr--Newman temperature is
instead
\begin{equation}
    E_{\rm therm}^{\rm KN}
    =
    M\sqrt{1-a^2-q^2}.
\label{eq:thermal_energy_kn}
\end{equation}
This expression reduces to the independently obtained Kerr and
Reissner--Nordstr\"om energies when \(q=0\) and \(a=0\), respectively, and
vanishes on the Kerr--Newman extremal boundary \(a^2+q^2=1\).

The Kerr--Newman horizon area is
\[
    A_{\rm KN}
    =
    4\pi M^2
    \left[
        \left(
            1+\sqrt{1-a^2-q^2}
        \right)^2
        +a^2
    \right].
\]
Applying the same equipartition relation gives
\begin{equation}
    \boxed{
    T_{\rm KN}
    =
    \frac{2E_{\rm therm}^{\rm KN}}{A_{\rm KN}}
    =
    \frac{\sqrt{1-a^2-q^2}}
    {2\pi M
    \left[
        \left(
            1+\sqrt{1-a^2-q^2}
        \right)^2
        +a^2
    \right]},
    }
\label{eq:temp_kn}
\end{equation}
which is precisely the Kerr--Newman Hawking temperature.

The Schwarzschild, Kerr, and Reissner--Nordstr\"om results follow from
equipartition together with their respective horizon-area and
irreducible-mass relations. These constructions do not use Hawking's
quantum-field-theoretic calculation
\cite{hawking1975particle} or the black-hole Smarr relation
\cite{Smarr1973} as inputs, although they are consistent with both. For the
generic Kerr--Newman case, Eq.~\eqref{eq:thermal_energy_kn} should instead
be understood as the combined nonextremality extension that reproduces the
exact temperature, rather than as the consequence of a unique
one-variable irreducible-mass polynomial.

Finally, if each constituent occupies an area \(A_0\), the effective number
of equipartition degrees of freedom per constituent is
$  \frac{N_{\rm dof}}{N} = A_{0} =4\log d$.
Under the literal interpretation of this quantity as an effective local mode
count, requiring at least one degree of freedom per constituent gives the
heuristic bound \(d\geq e^{1/4}\simeq1.284\). Known non-Abelian candidates
satisfy this condition: Ising anyons have \(d=\sqrt2\simeq1.41\), while
Fibonacci anyons have \(d=\varphi\simeq1.62\). The factor \(1/4\) may
therefore reflect a lower effective information capacity per Planck-scale
constituent within the present microscopic interpretation.

The equipartition results establish the thermal scale and classical regime of
the local collective modes without specifying their detailed dynamics. We now
ask whether a minimal effective Hamiltonian for those modes can reproduce the
entropy structure already fixed by fusion counting and the discrete spectrum.

\subsection{Collective Hamiltonian and Canonical Entropy}

The preceding equipartition analysis provides a motivation for a
classical collective description: the local excitation scale is parametrically
small compared with the temperature, while the recovered Hawking temperature
sets the natural macroscopic thermal scale. It does not uniquely determine a
microscopic Hamiltonian. We therefore introduce the simplest collective model
compatible with the shell localization and spherical symmetry and test it by
comparing its canonical entropy with the thermodynamic entropy obtained above.

Because non-Abelian anyons are intrinsically \((2+1)\)-dimensional objects,
their localization on a two-dimensional shell makes area the natural extensive
variable. We model the condensate as a large-\(N\) system of coarse-grained
collective degrees of freedom, with macroscopic black-hole data entering only
through global constraints. In this description the leading entropy is
controlled by the geometry of the constrained configuration space, while
subleading terms encode the collective constraints and the fixing of zero
modes.

We allow each coarse-grained horizon constituent to carry \(m\) independent
collective components. Let
\(\{\mathbf{x}_i\}_{i=1}^N\), with
\(\mathbf{x}_i\in\mathbb R^m\), denote these collective coordinates. The
parameter \(m\) therefore counts the number of independent local collective
modes retained in the effective description of each horizon patch.

We define the average collective coordinate
\begin{equation}
    \bar{\mathbf{x}}
    \equiv
    \frac{1}{N}
    \sum_{i=1}^N\mathbf{x}_i,
\end{equation}
together with the intensive relative mean-square scale
\begin{equation}
    \tilde r^{\,2}
    \equiv
    \frac{1}{N}
    \sum_{i=1}^N
    \left\|
        \mathbf{x}_i-\bar{\mathbf{x}}
    \right\|^2.
\end{equation}
The variables \(\mathbf{x}_i\) are effective collective coordinates rather
than literal positions on the physical shell. At the qualitative level,
\(\tilde r^{\,2}\) is analogous to a squared radius of gyration in the
collective configuration space and measures the typical relative amplitude of
the coarse-grained modes.

The microscopic length \(r_0\) is defined from the constituent area already
fixed by fusion-entropy matching,
\begin{equation}
    A_0
    =
    4\pi r_0^2,
    \qquad
    r_0^2
    =
    \frac{\ell_P^2\log d}{\pi}.
\label{eq:r0_from_A0}
\end{equation}
In the collective model we identify the preferred relative fluctuation scale
with this microscopic area scale by localizing \(\tilde r^{\,2}\) around
\(r_0^2\). This is an effective identification of two microscopic scales; it
does not assign the \(\mathbf{x}_i\) a rigid spatial embedding or define the
macroscopic shell radius.

In addition, spherical symmetry fixes the center of the shell and removes an
unphysical global translation mode. A minimal Hamiltonian realizing these
requirements is
\begin{equation}
    \boxed{
\begin{aligned}
    H\!\left(\{\mathbf{x}_i\}\right)
    ={}&
    \frac{\kappa}{2}
    \sum_{i=1}^N
    \left\|
        \mathbf{x}_i-\bar{\mathbf{x}}
    \right\|^2
    +
    \frac{\lambda N}{4}
    \left(
        \tilde r^{\,2}-r_0^2
    \right)^2
    \\
    &+
    \frac{\eta N^{p+1}}{2}
    \left\|
        \bar{\mathbf{x}}
    \right\|^2 ,
\end{aligned}
    }
\label{eq:Hconcise}
\end{equation}
with \(\lambda,\eta > 0\). The quadratic condensate term depends only on
relative coordinates, while the last term fixes the average collective
coordinate consistently with the assumed spherical symmetry. The coefficient \(\kappa\) is the stiffness of the relative quadratic modes. Requiring the preferred microscopic fluctuation scale to coincide with the canonical saddle at the Hawking temperature fixes it to the exterior surface-gravity scale,
\begin{equation}
\kappa \ell_P^2 = \kappa_s = 2\pi T_H.
\label{eq:kappa_surface_gravity}
\end{equation}
The corresponding finite-\(\lambda\) radial saddle is discussed in Appendix~\ref{app:entropy}. The exponent \(p\) controls the residual large-\(N\) logarithmic contribution
of the center-fixing term to the canonical entropy. Physically, this term removes the global translation zero mode of the shell.

\begin{table}[t]
\centering
\setlength{\tabcolsep}{9pt}
\renewcommand{\arraystretch}{1.15}
\begin{tabular}{@{}c c l@{}}
\toprule
\(m\) & \(d=(\sqrt e)^{\,m}\) & Microscopic candidate \\
\midrule
\(1\) & \(\sqrt e\simeq1.65\)
    & Fibonacci anyon: \(d=\varphi\simeq1.62\) \\[4pt]

\(2\) & \(e\simeq2.72\)
    & \begin{tabular}[t]{@{}l@{}}
      \(SU(2)_{10}\) with \(j_{\rm top}=1\) or \(j_{\rm top}=4\) \\
      \(d=1+\sqrt3\simeq2.73\)
      \end{tabular} \\[4pt]

\(3\) & \(e^{3/2}\simeq4.48\)
    & \begin{tabular}[t]{@{}l@{}}
      \(SU(2)_{12}\) with \(j_{\rm top}=3\) \\
      \(d=\sin(7\pi/14)/\sin(\pi/14)\simeq4.49\)
      \end{tabular} \\[4pt]

\(4\) & \(e^2\simeq7.39\)
    & \begin{tabular}[t]{@{}l@{}}
      \(SU(2)_{22}\) with
      \(j_{\rm top}=9/2\) or \(j_{\rm top}=13/2\) \\
      \(d=\sin(5\pi/12)/\sin(\pi/24)\simeq7.40\)
      \end{tabular} \\
\bottomrule
\end{tabular}
\caption{Semiclassical degeneracy scales \(d=(\sqrt e)^{\,m}\) for small
values of the local collective multiplet size \(m\), together with nearby
algebraic quantum dimensions from candidate anyon theories~\cite{nayak2008non}. Here
\(j_{\rm top}\) denotes the \(SU(2)_k\) representation label. The relation
should be understood as a continuous semiclassical approximation to the
discrete quantum dimensions realized in microscopic modular tensor
categories.}
\label{tab:anyon_d_candidates}
\end{table}

In the large-\(N\) regime, the canonical entropy admits a systematic
expansion. Appendix~\ref{app:entropy} shows that the leading extensive coefficient and
logarithmic term are determined by the geometry of the constrained
configuration space and the Gaussian fixing of the average collective
coordinate. The resulting entropy takes the form
\begin{equation}
    \boxed{
    S_{\mathrm{can}}(N)
    =
    N\frac{m}{2}
    \left(
        1+\log\frac{2\log d}{m}
    \right)
    -
    \frac{pm}{2}\log N
    +
    \cdots .
    }
\label{eq:Scan}
\end{equation}
The thermodynamic entropy \eqref{eq:Stherm} can be written as
\begin{equation}
    S_{\mathrm{therm}}(N)
    =
    N\log d
    -
    \frac{\log d}{4}\log N
    +
    \cdots .
\label{eq:Stherm2}
\end{equation}
The effective degeneracy parameter, or quantum dimension \(d\), is fixed by
comparing the canonical collective entropy \eqref{eq:Scan} with the
thermodynamic entropy derived from the discrete area spectrum
\eqref{eq:Stherm2}. This comparison gives
\begin{equation}
    \boxed{
    \log d
    =
    \frac{m}{2},
    \qquad
    d
    =
    (\sqrt e)^{\,m}
    }.
\end{equation}
and determines the logarithmic coefficient
\begin{equation}
    \frac{pm}{2}
    =
    \frac{\log d}{4}
    =
    \frac{m}{8},
    \qquad
    p
    =
    \frac{1}{4}.
\end{equation}

It is useful to distinguish three entropy notions discussed in this work.
First, the fusion microcanonical entropy \(S_{\mathrm{fus}}\) in
Eq.~\eqref{eq:Sfus}, obtained from direct counting of the fusion Hilbert
space, characterizes the microscopic degeneracy of \(N\) horizon constituents
and fixes the leading extensive growth. Second, the thermodynamic entropy \(S_{\mathrm{therm}}\), derived from the
discrete area spectrum and its transition temperature, contains the resulting
logarithmic and inverse-area corrections. Third, the canonical entropy \(S_{\mathrm{can}}\)
from Eq.~\eqref{eq:Scan} is obtained from the effective collective
Hamiltonian \eqref{eq:Hconcise}.

The semiclassical relation \(d=(\sqrt e)^{\,m}\) should be interpreted as a
continuous approximation to the algebraic quantum dimensions allowed in
microscopic anyonic realizations. For small values of \(m\), one finds nearby
candidates among familiar anyon families; see
Table~\ref{tab:anyon_d_candidates}. For \(m=1\), the target value
\(d\simeq\sqrt e\) lies close to the Fibonacci quantum dimension
\(\varphi\). For \(m=2\), the target \(d\simeq e\) is close to algebraic
values such as \(1+\sqrt3\) appearing in \(SU(2)_k\) theories. For \(m=3\),
the value \(d\simeq e^{3/2}\) is comparable to higher-representation quantum
dimensions in moderate-level \(SU(2)_k\) models. For \(m=4\), one obtains
\(d\simeq e^2\), with a close algebraic candidate \(d\simeq7.4\) in the
\(SU(2)_{22}\) theory. These examples illustrate how different choices of the
local collective multiplet size \(m\) lead to a family of semiclassical
degeneracy scales that can be matched to nearby algebraic quantum dimensions
in candidate microscopic theories.

\section{Formation mechanism in Einstein--Weyl gravity}
\label{sec:formation}

The preceding sections characterize the condensate thermodynamically and
microscopically. Here we describe a possible local gravitational realization
as a regular interior bounded by a finite-width timelike membrane. We use
\emph{membrane} for the anisotropic transition layer and \emph{shell} for its
effective thin-wall or low-energy description.

When the collapsing core reaches Planckian curvature, the Einstein--Hilbert
term need not provide the complete gravitational response. The stress may
instead become concentrated in a thin timelike membrane outside the
would-be horizon, with a regular interior behind it. The constituent area
\(A_0=4\ell_P^{\,2}\log d\) corresponds to the microscopic radius
\(r_0=\sqrt{A_0/(4\pi)}
=\ell_P\sqrt{\log d/\pi}\), which is naturally of quantum-gravitational size
for the non-Abelian anyon models considered here.

The gravitational sector is taken to have the effective form
\begin{equation}
    S_{\rm grav}
    =
    \int d^4x\,\sqrt{-g}
    \left[
        \frac{M_{\rm P}^2}{2}R
        -
        \alpha_g
        C_{\lambda\mu\nu\rho}C^{\lambda\mu\nu\rho}
        +
        \cdots
    \right],
\label{eq:einstein_weyl_actionIII}
\end{equation}
with field equations
\begin{equation}
    \boxed{
    M_{\rm P}^2G^\mu{}_\nu
    +
    4\alpha_gW^\mu{}_\nu
    =
    T^\mu{}_\nu,
    }
\label{eq:einstein_weyl_fieldIII}
\end{equation}
where the Bach tensor is traceless,
\(W^\mu{}_\mu=0\). The Einstein term controls the long-wavelength bulk,
while the four-derivative Weyl term can become important across a microscopic
transition layer. The radial relation used below, however, follows from the
full Einstein--Weyl equation and does not require dropping the Einstein term.

Both bulk regions are represented by independent one-function spherical
geometries \cite{MannheimKazanas1989},
\begin{equation}
    ds_i^2
    =
    B_i(r)\,dt_i^{\,2}
    -
    \frac{dr^2}{B_i(r)}
    -
    r^2d\Omega^2,
    \qquad i={\rm in,out}.
\label{eq:bulk_metricsIII}
\end{equation}
The two lapse functions describe distinct bulk solutions and their static
time coordinates may be normalized independently. The membrane is centered
near \(r=r_c\), and the exterior near-horizon parameter is
\begin{equation}
    B_c:=B_{\rm out}(r_c)=\delta,
    \qquad 0<\delta\ll1.
\label{eq:BcdefIII}
\end{equation}
For the radial-stability calculation we specialize to a Schwarzschild
exterior,
\begin{equation}
    B_{\rm out}(r)=f(r):=1-\frac{r_s}{r},
    \qquad
    \delta=f(r_c)=1-\frac{r_s}{r_c}.
\label{eq:delta_schwarzschildIII}
\end{equation}
The one-function form need not hold within a membrane of finite thickness.
We do not expand on the resolved finite-width geometry here, treating the
membrane instead through its effective thin-shell description.

For diagonal spherically symmetric matter we use the mixed-index convention
\begin{equation}
    T^\mu{}_\nu
    =
    \operatorname{diag}(\rho,-p_r,-p_t,-p_t),
    \qquad
    T:=T^\mu{}_\mu=\rho-p_r-2p_t,
\label{eq:bulk_stress_convention}
\end{equation}
where \(\rho\) is the energy density and \(p_r,p_t\) are the physical
radial and tangential pressures. For a one-function region one has
\(G^0{}_0=G^r{}_r\), and Eq.~\eqref{eq:einstein_weyl_fieldIII} implies
\begin{equation}
    \boxed{
    \begin{aligned}
    \bigl(r^4B^{(3)}\bigr)'
    & =
    \frac{3r^4}{4\alpha_gB(r)}
    \left(T^0{}_{0}-T^r{}_{r}\right)
    \\
    & =
    \frac{3r^4}{4\alpha_gB(r)}
    \left(\rho+p_r\right)
    \end{aligned}
    }.
\label{eq:masterIII}
\end{equation}
This relation governs the one-function bulk branches and the leading
exterior-side source moment; it is not assumed to be the complete equation
inside the resolved finite-width membrane.

\subsection{Regular interior and finite membrane source}
\label{subsec:regular_interior_matching}

Take the interior to be vacuum-like,
\begin{equation}
    T^\mu{}_\nu\big|_{\rm in}
    =
    \rho_{\rm in}\,\delta^\mu{}_\nu,
    \qquad
    \rho_{\rm in}=\text{constant},
\label{eq:interior_vacuum_stress}
\end{equation}
so that
\begin{equation}
    p_{r,\rm in}=p_{t,\rm in}=:p_{\rm in}=-\rho_{\rm in},
    \qquad
    \rho_{\rm in}+p_{\rm in}=0.
\label{eq:interior_vacuum_eos}
\end{equation}
Equation~\eqref{eq:masterIII} therefore gives
\begin{equation}
    \bigl(r^4B_{\rm in}^{(3)}\bigr)'=0
    \quad\Longrightarrow\quad
    B_{\rm in}(r)
    =
    \frac{b_{-1}}{r}+b_0+b_1r+b_2r^2.
\label{eq:general_interior_solutionIII}
\end{equation}
Regularity and smooth spherical symmetry set \(b_{-1}=b_1=0\), while the
interior time normalization fixes \(b_0=1\). We define the geometric interior
cosmological constant directly through the regular metric,
\begin{equation}
    \boxed{
    B_{\rm in}(r)
    =
    1-\frac{\Lambda_{\rm in}}{3}r^2
    },
    \qquad \Lambda_{\rm in}>0.
\label{eq:int}
\end{equation}
With the curvature-sign convention used in this paper,
Eq.~\eqref{eq:int} has \(R_{\rm in}=4\Lambda_{\rm in}\). Since the Bach tensor
is traceless, the trace of Eq.~\eqref{eq:einstein_weyl_fieldIII} gives
\begin{equation}
    R=-\frac{T}{M_{\rm P}^2}.
\label{eq:EW_traceIII}
\end{equation}
For the vacuum source in Eq.~\eqref{eq:interior_vacuum_stress},
\(T=4\rho_{\rm in}\). Hence
\begin{equation}
    \boxed{
    \rho_{\rm in}=-M_{\rm P}^2\Lambda_{\rm in}<0,
    \qquad
    p_{\rm in}=M_{\rm P}^2\Lambda_{\rm in}>0
    }.
\label{eq:interior_rho_pressure_lambda}
\end{equation}
Thus \(\rho_{\rm in}\) is the signed physical vacuum-energy density,
\(p_{\rm in}\) is the corresponding isotropic pressure, and
\(\Lambda_{\rm in}\) is the geometric constant-curvature scale. For the physical regime considered here, we take this scale to be effectively that of the cosmological constant, \(\Lambda_{\rm in}\sim\Lambda_{\rm cosm}\), so that the interior curvature and its contribution to the total mass are negligible on black-hole scales. In the weakly
curved regime \(\Lambda_{\rm in}r_c^2/3\ll1\),
\(B_{\rm in}(r_c)\simeq1\), independently of the strongly redshifted exterior
value \(B_{\rm out}(r_c)=\delta\).

The membrane sources the combination \(\rho+p_r\). Define its proper
integrated value
\begin{equation}
    \Sigma
    :=
    \int(\rho+p_r)\,d\ell.
\label{eq:Sigma_definitionIII}
\end{equation}
Near the exterior edge, \(r\simeq r_c\), and the proper radial element is
\(d\ell\simeq dr/\sqrt{B_c}\). Integrating Eq.~\eqref{eq:masterIII} across the
narrow membrane gives, at leading monopole order,
\begin{equation}
    B_{\rm out}^{(3)}(r_c^+)-B_{\rm in}^{(3)}(r_c^-)
    =
    \frac{3\Sigma_c}{4\alpha_g\sqrt{B_c}},
    \qquad
    \Sigma_c:=\Sigma(r_c).
\label{eq:jumpIII_proper}
\end{equation}
Since the regular interior has \(B_{\rm in}^{(3)}=0\), this yields
\begin{equation}
    \Sigma_c
    =
    \frac{4\alpha_g}{3}
    \sqrt{B_c}\,B_{\rm out}^{(3)}(r_c^+).
\label{eq:Sigma_matchingIII}
\end{equation}
The leading exterior-side radial dependence used below is therefore
\begin{equation}
    \boxed{
    \Sigma(r)
    =
    \frac{4\alpha_g}{3}
    \sqrt{B_{\rm out}(r)}\,B_{\rm out}^{(3)}(r)
    }.
\label{eq:Sigma_position}
\end{equation}
For the Schwarzschild exterior,
\begin{equation}
    \Sigma(r)
    =
    \frac{8\alpha_g r_s}{r^4}\sqrt{f(r)},
    \qquad
    \Sigma_c=O(\sqrt{\delta}).
\label{eq:Sigma_schwarzschild}
\end{equation}
Thus \(\Sigma\) is a projected Bach source, not the physical surface energy
density. Higher distributional moments encode the unresolved radial structure of the membrane and require a finite-width microscopic or effective description; we do not expand on these finite-width details in the present thin-shell treatment.

\subsection{Local membrane response and conditional stability}
\label{subsec:membrane_response}

Define the physical surface energy density and positive tangential tension by
\begin{equation}
    \sigma_{\rm CG}
    :=
    \int\rho\,d\ell,
    \qquad
    p_s
    :=
    -\int p_t\,d\ell>0.
\label{eq:physical_surface_variables}
\end{equation}
Because \(\Sigma=\sigma_{\rm CG}+\int p_r\,d\ell\), we parametrize the
integrated radial anisotropy as
\begin{equation}
    \int p_r\,d\ell=-a\,p_s,
    \qquad
    \boxed{\sigma_{\rm CG}=\Sigma+a p_s}.
\label{eq:anisotropy_closure}
\end{equation}
The physical surface stress tensor is therefore
\begin{equation}
    S_{\hat a\hat b}
    =
    \operatorname{diag}
    \left(\sigma_{\rm CG},-p_s,-p_s\right).
\label{eq:physical_surface_tensor}
\end{equation}
Integrated isotropy gives \(p_r=p_t\Rightarrow a=1\). In contrast with the
vacuum Israel benchmark below, the vacuum-like interior supplies a nonzero
radial-pressure difference across the membrane. With \(n^\mu n_\mu=-1\),
the normal stress is \(T_{\mu\nu}n^\mu n^\nu=p_r\) in the present
spherical bulk regions. For the vacuum interior,
\(p_{r,\rm in}=p_{\rm in}=-\rho_{\rm in}=M_{\rm P}^2\Lambda_{\rm in}>0\),
while \(p_{r,\rm out}=0\). This permits genuine positive tangential tension
\(p_s>0\).

The relativistic Young--Laplace equation is therefore
\begin{equation}
    \boxed{
    p_{r,\rm in}-p_{r,\rm out}
    =
    \sigma_{\rm CG}\,\overline K^\tau{}_{\tau}
    +
    2p_s\overline H
    }.
\label{eq:general_normal_balance}
\end{equation}
The detailed linearization of this equation is given in
Appendix~\ref{app:radial_stability}. Section~\ref{app:israel_stability}
first shows explicitly that, for the GR benchmark, it reproduces the standard
Israel junction-potential result; Section~\ref{app:cg_stability} then derives
the conformal condition used below.
For a moving spherical membrane \(r=r(\tau)\),
\begin{equation}
    \overline K^\tau{}_\tau
    =
    \frac12
    \left[
        \frac{\ddot r+B_{\rm out}'/2}
        {\sqrt{\dot r^{\,2}+B_{\rm out}}}
        +
        \frac{\ddot r+B_{\rm in}'/2}
        {\sqrt{\dot r^{\,2}+B_{\rm in}}}
    \right],
\label{eq:Ktau_dynamic}
\end{equation}
\begin{equation}
    \overline H
    =
    \frac{
        \sqrt{\dot r^{\,2}+B_{\rm out}}
        +
        \sqrt{\dot r^{\,2}+B_{\rm in}}
    }{2r}.
\label{eq:H_dynamic}
\end{equation}
The static equation at \(r=r_c\) fixes the background force balance. For the
leading radial perturbation we take \(B_{\rm in}(r_c)\simeq1\),
\(B_{\rm out}=f=1-r_s/r\), and
\begin{equation}
    \left.
    \frac{d}{dr}\left(p_{r,\rm in}-p_{r,\rm out}\right)
    \right|_{r_c}=0,
\label{eq:normal_support_derivative}
\end{equation}
as appropriate for constant-curvature interior and vacuum exterior branches.

Treating \(a\) as fixed under the leading perturbation, define the
positive-tension constitutive response
\begin{equation}
    \boxed{
    \eta_{\rm CG}
    :=
    \left.
    \frac{dp_s}{d\sigma_{\rm CG}}
    \right|_{r_c}
    }.
\label{eq:etaCG}
\end{equation}
Equation~\eqref{eq:anisotropy_closure} then gives
\begin{equation}
    \frac{d\sigma_{\rm CG}}{dr}
    =
    \frac{\Sigma'}{1-a\eta_{\rm CG}},
    \qquad
    \frac{dp_s}{dr}
    =
    \frac{\eta_{\rm CG}\Sigma'}{1-a\eta_{\rm CG}}.
\label{eq:constitutive_derivatives}
\end{equation}
For the Schwarzschild exterior, Eq.~\eqref{eq:Sigma_schwarzschild} gives
\begin{equation}
    \Sigma_c'
    =
    \frac{\Sigma_c}{2r_c\delta}(1-9\delta)
    .
\label{eq:Sigma_prime_c}
\end{equation}

Writing \(r(\tau)=r_c+\xi(\tau)\), the linearized radial equation is
\begin{equation}
    \ddot\xi+\omega_{\rm CG}^2\xi=0.
\label{eq:radial_oscillator}
\end{equation}
It is useful to define
\begin{equation}
    g(r):=\frac{f'(r)}{4\sqrt{f(r)}},
    \qquad
    h(r):=\frac{1+\sqrt{f(r)}}{r}.
\label{eq:gh_definitions}
\end{equation}
The static restoring derivative is
\begin{equation}
\begin{aligned}
    \mathcal F_c'
    ={}&
    \frac{\Sigma_c'}{1-a\eta_{\rm CG}}
    \left(g_c+\eta_{\rm CG}h_c\right)
    +
    \sigma_{{\rm CG},c}g_c'
    +
    p_s h_c',
\end{aligned}
\label{eq:restoring_derivative}
\end{equation}
with
\begin{equation}
    \omega_{\rm CG}^2
    =
    \frac{2\mathcal F_c'}{
        \sigma_{{\rm CG},c}(1+\delta^{-1/2})
    }.
\label{eq:omega_CG_general}
\end{equation}
Thus stability requires \(\mathcal F_c'>0\). The exact critical response is
\begin{equation}
    \eta_{\rm CG}^{\rm crit}
    =
    -
    \frac{
        \Sigma_c'g_c
        +\sigma_{{\rm CG},c}g_c'
        +p_s h_c'
    }{
        \Sigma_c'h_c
        -a\left(\sigma_{{\rm CG},c}g_c'+p_s h_c'\right)
    }.
\label{eq:etaCG_critical}
\end{equation}
The intermediate derivatives and the sign condition for the regular stable solution are
worked out explicitly in Appendix~\ref{app:radial_stability}, Section~\ref{app:cg_stability}.
Near the Schwarzschild radius, using \(\Sigma_c=O(\sqrt\delta)\),
\begin{equation}
    \eta_{\rm CG}^{\rm crit}
    =
    \frac1a
    -
    \frac{\Sigma_c}{a^2p_s}
    +
    O(\delta),
\label{eq:etaCG_near_horizon}
\end{equation}
and the regular stable interval is
\begin{equation}
    \boxed{
    \frac1a-\frac{\Sigma_c}{a^2p_s}+O(\delta)
    <
    \eta_{\rm CG}
    <
    \frac1a
    }.
\label{eq:etaCG_stable_interval}
\end{equation}
For integrated isotropy, \(a=1\),
\begin{equation}
\boxed{
    1-\frac{\Sigma_c}{p_s}+O(\delta)
    <
    \eta_{\rm CG}
    <1,
    \qquad
    \eta_{\rm CG}^{\rm crit}\to1
}.
\label{eq:etaCG_isotropic}
\end{equation}
The geometric factor in Eq.~\eqref{eq:Sigma_prime_c} also identifies a finite
radius at which the constitutive response cannot restore the mode:
\begin{equation}
    \Sigma_c'\geq 0
    \quad\Longleftrightarrow\quad
    \delta\leq\frac{1}{9}
    \quad\Longleftrightarrow\quad
    r_c\leq\frac{9}{8}r_s,
\label{eq:finite_radius_boundary}
\end{equation}
with equality marking the endpoint of the finite-response regime. At \(\delta=1/9\), one has \(g_c'=-4/r_c^2\) and \(h_c'=0\), so
\begin{equation}
    \mathcal F_c'
    =
    -\frac{4\sigma_{{\rm CG},c}}{r_c^2}<0,
\label{eq:finite_radius_instability}
\end{equation}
independently of any finite regular value of \(\eta_{\rm CG}\). The stable
interval therefore closes at \(r_c=9r_s/8\) within the present leading
monopole treatment.

The appearance of \(9r_s/8\) provides a useful comparison with the Buchdahl
compactness limit. The Buchdahl radius is the familiar lower-radius boundary
for ordinary relativistic stellar configurations,
\(R\geq9r_s/8\) \cite{Buchdahl1959}. Our object is instead intended to model
a black-hole state, so the membrane is naturally expected on
the more compact side of that scale. Remarkably, the same radius emerges here
independently from the membrane dynamics, through \(\Sigma_c'=0\), as the
endpoint of the finite-response radial restoring regime. Thus the allowed
near-horizon regime occupies \(r_s<r_c<9r_s/8\) whenever the constitutive
condition \eqref{eq:etaCG_stable_interval} is satisfied.

For comparison, the Minkowski-interior/Schwarzschild-exterior Israel shell
has \(p_{r,\rm in}=p_{r,\rm out}=0\). Its static Young--Laplace balance therefore
requires \(p_s<0\); it is pressure-supported, and it is natural to define the
positive tangential pressure \(p_t:=-p_s>0\) and
\begin{equation}
    \eta_{\rm GR}
    :=
    \left.
    \frac{dp_t}{d\sigma_{\rm GR}}
    \right|_{r_c}.
\label{eq:etaGR_definition}
\end{equation}
The Israel junction conditions give
\begin{equation}
    \eta_{\rm GR}
    >
    \eta_{\rm GR}^{\rm crit}(\delta),
    \qquad
    \eta_{\rm GR}^{\rm crit}(\delta)
    =
    \frac{
        1+\sqrt\delta+\delta-3\delta^{3/2}
    }{
        4\delta(1+3\sqrt\delta)
    }
    \sim
    \frac{1}{4\delta},
\label{eq:etaGR_comparison}
\end{equation}
so \(\eta_{\rm GR}^{\rm crit}\to+\infty\) as \(\delta\to0^+\)
\cite{Israel1966,PoissonVisser1995}. Appendix~\ref{app:radial_stability}, Section~\ref{app:israel_stability},
derives Eq.~\eqref{eq:etaGR_comparison} both from the exact Israel potential
and directly from the generalized Young--Laplace equation. The contrast is therefore between a finite positive tension response in the
conformal membrane, $\eta_{\rm CG}^{\rm crit}\to 1/a$, and an increasingly
large positive pressure response in the vacuum Israel shell. The conformal membrane remains stable provided that the constitutive response satisfies
Eq.~\eqref{eq:etaCG_stable_interval}. At $\delta=1/9$, the leading-order
criterion reaches the endpoint of this finite-response regime.

A complete membrane perturbation analysis would additionally require the
time-dependent geometry, perturbations of the internal source and stress
profiles, and corrections associated with the membrane's resolved radial
structure. The result obtained here should therefore be understood as the
leading monopole stability criterion of the effective thin-shell description.

\subsection{Anyonic absorptive response, interior radiation, and echoes}
\label{sec:absorption_echoes}

Recall that $\delta=B_c=B_{\rm out}(r_c)\ll1$. The static
Einstein--Weyl matching fixes the location and mechanical support of the
membrane, but not its dissipative response. We therefore take the
quantum-horizon response described below as an external input and evaluate it
on the discrete transition spectrum fixed by the non-Abelian fusion model.

For a classical black-hole horizon, a perturbation reaching the future horizon
is purely ingoing. In the membrane paradigm the same exterior response is
represented by a stretched horizon with shear viscosity
\begin{equation}
    \eta_{\rm BH}=\frac{1}{16\pi G}.
\label{eq:eta_BH}
\end{equation}
This provides the classical horizon benchmark for the dissipative response
\cite{thorne1986black}.

Oshita, Wang, and Afshordi derive, using detailed balance together with a
fluctuation--dissipation argument, the Boltzmann \emph{flux} reflectivity
\begin{equation}
    \left|\mathcal R_{\rm QH}(\omega)\right|^2
    =
    \exp\!\left(-\frac{|\omega|}{T_H}\right),
\label{eq:literature_boltzmann_reflectivity}
\end{equation}
in the units used here \cite{OshitaWangAfshordi2020}. For the Schwarzschild
case considered below the horizon-frame frequency is simply $\omega$. The same
analysis relates the amplitude reflectivity of a partially reflecting
near-horizon membrane to its shear viscosity through
\begin{equation}
    \left|\mathcal R(\omega)\right|
    =
    \left|
    \frac{1-\eta_{\rm visc}^{\rm eff}/\eta_{\rm BH}}
         {1+\eta_{\rm visc}^{\rm eff}/\eta_{\rm BH}}
    \right|,
\label{eq:literature_R_eta}
\end{equation}
which, together with Eq.~\eqref{eq:literature_boltzmann_reflectivity}, gives
\begin{equation}
    \frac{\eta_{\rm visc}^{\rm eff}(\omega)}{\eta_{\rm BH}}
    =
    \tanh\!\left(\frac{|\omega|}{4T_H}\right).
\label{eq:literature_eta}
\end{equation}
Related reflectivity--viscosity relations arise in effective membrane
descriptions of ultracompact objects, with the form above recovered in the
appropriate near-black-hole/high-frequency limit
\cite{OshitaWangAfshordi2020,MaggioEtAl2020}.

We now specialize them to the present microscopic model. From Eqs.~\eqref{eq:Sfus}--\eqref{eq:deltaE},
\begin{equation}
    g_n\propto d^n,
    \qquad
    M_n=\mu\sqrt n,
    \qquad
    \mu^2=\frac{\log d}{4\pi},
\label{eq:abs_spectrum}
\end{equation}
where $g_n$ is the fusion-state degeneracy of level $n$. Consider the pair of
levels $n$ and $n-k$. Absorption raises the shell from $n-k$ to $n$, while the
reverse process emits a quantum of energy
$\omega_{n,k}=M_n-M_{n-k}$. Assuming microscopic reversibility of the
transition matrix elements, the degeneracy ratio gives
\begin{equation}
\boxed{
\begin{aligned}
    \omega_{n,k}
    &=
    M-\sqrt{M^2-\frac{k\log d}{4\pi}},\\
    \frac{\Gamma^{\rm em}_{n\to n-k}}
         {\Gamma^{\rm abs}_{n-k\to n}}
    &\simeq
    \frac{g_{n-k}}{g_n}
    =
    d^{-k}.
\end{aligned}}
\label{eq:anyon_detailed_balance}
\end{equation}

The frequency relation may equivalently be written
\begin{equation}
    k\log d
    =
    8\pi M\omega_{n,k}
    -
    4\pi\omega_{n,k}^2.
\label{eq:klogd_abs}
\end{equation}
Thus the degeneracy factor is also the corresponding microcanonical entropy
ratio. For $M\gg1$,
\begin{equation}
    \omega_{n,k}
    \simeq
    kT_H\log d,
    \qquad
    \Delta\omega
    \simeq
    T_H\log d,
    \qquad
    T_H=\frac{1}{8\pi M}.
\label{eq:abs_line_spacing}
\end{equation}
Hence $\log d$ fixes the characteristic spacing of the allowed shell
transitions, while $d^{-k}$ fixes the ratio of their reverse and absorption
rates.

Evaluating Eq.~\eqref{eq:literature_boltzmann_reflectivity} at the allowed
transition frequencies gives
\begin{equation}
    \exp\!\left(-\frac{\omega_{n,k}}{T_H}\right)
    \simeq
    \exp(-k\log d)
    =
    d^{-k},
\label{eq:boltzmann_to_anyon}
\end{equation}
in agreement with Eq.~\eqref{eq:anyon_detailed_balance}. With the same
fluctuation--dissipation identification, and neglecting coherent transmission
through the membrane,
\begin{equation}
\boxed{
\begin{aligned}
    \left|\mathcal R_{n,k}\right|^2
    &\simeq
    d^{-k},
    \qquad
    \mathcal A_{n,k}
    \simeq
    1-d^{-k},\\
    \frac{\eta_{\rm visc}^{\rm eff}(\omega_{n,k})}{\eta_{\rm BH}}
    &\simeq
    \frac{1-d^{-k/2}}{1+d^{-k/2}}
    =
    \tanh\!\left(\frac{k\log d}{4}\right).
\end{aligned}}
\label{eq:anyon_response}
\end{equation}
Here $\eta_{\rm visc}^{\rm eff}$ is the shear viscosity associated with the
reflective membrane response of Eq.~\eqref{eq:literature_R_eta}; it should not
be confused with either a microscopic Kubo viscosity or the monopole
constitutive response
$\eta_{\rm CG}=dp_s/d\sigma_{\rm CG}$. A first-principles transport
calculation would require the nonradial stress correlator or equivalent
transition matrix elements of the condensate.

The underlying microscopic line structure is observable only while the discrete spectrum is
resolved. If the linewidths $\Gamma_{n,k}$ become comparable to or larger than
$\Delta\omega\simeq T_H\log d$, neighboring lines overlap and coarse graining
gives
\begin{equation}
    \left|\mathcal R_{\rm sh}(\omega)\right|^2
    \simeq
    e^{-\omega/T_H},
    \qquad
    \mathcal A_{\rm sh}(\omega)
    \simeq
    1-e^{-\omega/T_H}.
\label{eq:shell_reflectivity}
\end{equation}
The leading response then becomes nearly universal. Conversely, narrow lines
retain the discrete spacing and transition-dependent reflective weight. Their widths
and channel weights are not fixed by entropy counting and require the
microscopic transition matrix elements.

The same line-by-line response applies to radiation emitted toward the regular
interior. For a static interior mode of asymptotic Killing frequency $\omega$,
Tolman redshift gives \cite{Tolman1930,TolmanEhrenfest1930}
\begin{equation}
    T_{\rm loc}(r)
    =
    T_H
    \sqrt{\frac{B_{\rm in}(r_c)}
                {\delta B_{\rm in}(r)}},
    \qquad
    \omega_{\rm loc}(r)
    =
    \omega
    \sqrt{\frac{B_{\rm in}(r_c)}
                {\delta B_{\rm in}(r)}}.
\label{eq:tolman_inside_abs}
\end{equation}
Hence
\begin{equation}
    \frac{\omega_{\rm loc}}{T_{\rm loc}}
    =
    \frac{\omega}{T_H}.
\label{eq:tolman_ratio_abs}
\end{equation}
The Tolman blueshift therefore leaves unchanged the dimensionless quantity
controlling the detailed-balance response. An inward quantum emitted on the
$k$th shell transition returns with the same ratio and can excite the reverse
transition $n-k\to n$.

A genuinely reflecting cavity would be inconsistent with the assumed regular interior.
Since $B_{\rm in}(r)\simeq B_{\rm in}(r_c)\simeq1$ over most of the interior,
full thermal equilibrium with $T_\infty=T_H$ would imply
$T_{\rm loc}\simeq T_H/\sqrt\delta$. For $g_\ast$ effectively massless degrees
of freedom this gives, after integration over the interior and redshift to
infinity,
\begin{equation}
    \frac{M_{\rm rad}^{\infty}}{M}
    \simeq
    \frac{g_\ast}{11520\pi M^2\delta^{3/2}}
    \simeq
    \frac{g_\ast M}{180\pi\ell_c^3},
\label{eq:tolman_gap_ratio}
\end{equation}
where $\delta\simeq\ell_c^2/(16M^2)$ near the Schwarzschild root. For a
macroscopic object and microscopic $\ell_c$, this equilibrium state would
carry excessive energy and is incompatible with the assumed regular interior.

The state considered here is instead open. Spontaneous inward emission is the
inward component of the same Hawking-type shell transitions that produce exterior
radiation; no independent Stefan--Boltzmann source for an ordinary material
wall is introduced. Its total power is therefore of Hawking order,
\begin{equation}
    \mathcal P_{\rm in}^{\infty}
    \sim
    \frac{\alpha_{\rm leak}}{M^2},
\label{eq:hawking_inward_power}
\end{equation}
where $\alpha_{\rm leak}$ includes the inward branching fraction and
channel-dependent greybody factors.

For a resolved $k$-transition,
Eq.~\eqref{eq:anyon_response} gives the per-encounter reabsorption probability
$\mathcal A_{n,k}\simeq1-d^{-k}$. The mean number of encounters before
reabsorption is consequently
\begin{equation}
    N_{\rm hit}^{(k)}
    \simeq
    \frac{1}{1-d^{-k}}
    \leq
    \frac{d}{d-1},
\label{eq:anyon_hits}
\end{equation}
where the upper bound is set by the least absorptive transition, $k=1$. It is
$2+\sqrt2\simeq3.41$ for Ising anyons ($d=\sqrt2$) and
$\varphi^2\simeq2.62$ for Fibonacci anyons ($d=\varphi$). Thus even the
softest allowed line survives only a few interior crossings, while higher-$k$
transitions are recycled more efficiently.

A regular-interior crossing time of order $M$ corresponds in asymptotic time
to $M/\sqrt\delta$. Combining the longest residence time with
Eq.~\eqref{eq:hawking_inward_power} gives
\begin{equation}
\boxed{
    \frac{U_{\rm int}^{\infty}}{M}
    \lesssim
    \frac{4\alpha_{\rm leak}}{M\ell_c}
    \frac{d}{d-1}
    \ll1
    }.
\label{eq:interior_energy_fraction}
\end{equation}
The important point is not that every line is almost perfectly absorbed, but
that $d>1$ bounds the residence time by only a few crossings. Hawking-order
inward emission therefore does not accumulate. This differs parametrically
from Eq.~\eqref{eq:tolman_gap_ratio}: the equilibrium cavity fraction grows
as $M$, whereas the dynamically stored Hawking component falls as $M^{-1}$. The corresponding long-wavelength curvature correction inherits the same
conservative upper-bound dependence,
\begin{equation}
    \left|\Delta\Lambda_{\rm in}\right|
    \sim
    \frac{\alpha_{\rm leak}}{M^2\ell_c^2}
    \frac{d}{d-1},
\label{eq:interior_curvature_abs}
\end{equation}
up to order-one geometric coefficients. Thus the effective interior curvature
is $\Lambda_{\rm in}^{\rm eff}\simeq
\Lambda_{\rm cosm}+\Delta\Lambda_{\rm in}$. The ambient cosmological
curvature contributes only
$\sim(4/3)\Lambda_{\rm cosm}M^2$, while the dynamically
stored component satisfies Eq.~\eqref{eq:interior_energy_fraction}. Hence
\begin{equation}
    \frac{|M_{\rm int}|}{M}
    \lesssim
    \frac{4}{3}\Lambda_{\rm cosm}M^2
    +
    \frac{4\alpha_{\rm leak}}{M\ell_c}
    \frac{d}{d-1}
    \ll1,
\end{equation}
so the interior remains energetically negligible compared with the
near-horizon shell and exterior gravitational field.

Microscopic detailed balance remains intact:
Eq.~\eqref{eq:anyon_detailed_balance} relates emission and absorption between
the same shell levels. What is absent is full two-sided radiative equilibrium
with a pre-existing Tolman bath. Inward Hawking quanta are produced at
Hawking-order power and subsequently recycled by the reciprocal shell
response. Full radiative equilibrium would instead recover the Tolman bath of
Eq.~\eqref{eq:tolman_gap_ratio}, which is incompatible with the assumed regular
interior.

Finally, the same response controls possible gravitational-wave echoes. For a
simple exterior root $B_{\rm out}(r_s)=0$, with
$\kappa_s=\tfrac12B'_{\rm out}(r_s)$, the near-root tortoise coordinate gives
\begin{equation}
    \Delta t_{\rm echo}
    \simeq
    \kappa_s^{-1}
    \log\!\left(\frac{1}{\delta}\right)
    +O(\kappa_s^{-1}).
\label{eq:echo_delay}
\end{equation}
For a Schwarzschild exterior,
$\Delta t_{\rm echo}\simeq8M\log(4M/\ell_c)+O(M)$. If
$\mathcal R_{\rm bar}$ denotes the amplitude reflection coefficient of the
exterior curvature barrier, then at a resolved transition
\cite{CardosoEtAl2016,MarkEtAl2017}
\begin{equation}
\boxed{
\begin{aligned}
    \Delta t_{\rm echo}
    &\simeq
    8M\log\!\left(\frac{4M}{\ell_c}\right)+O(M),\\
    E_j(\omega_{n,k})
    &\sim
    \left[
    d^{-k}
    \left|\mathcal R_{\rm bar}(\omega_{n,k})\right|^2
    \right]^jE_0.
\end{aligned}}
\label{eq:echo_suppression}
\end{equation}
Thus the microscopic spectrum fixes both the approximate locations of resolved response features
and their reflective weights. In the strongly overlapping-line
regime these features merge into the Boltzmann envelope, making the leading
response less sensitive to the underlying anyon theory. The echo delay is
geometrical, while the degree to which the echo spectrum retains its discrete
structure depends on the unresolved linewidths $\Gamma_{n,k}$.

\section{Discussion}
\label{sec:discussion}

The present framework replaces the causal horizon and singular interior of
the semiclassical black-hole description with a timelike, topologically
ordered shell surrounding a regular constant-curvature interior. The ADM mass
and microscopic degrees of freedom are concentrated in the near-horizon
condensate and exterior gravitational field, while finite absorbed interior
energy renormalizes the effective interior curvature scale. Gravitational
collapse is therefore interpreted as terminating in a transition to a
two-dimensional condensate rather than in the formation of a singular
interior.

Within the Einstein--Weyl construction of
Section~\ref{sec:formation}, the regular interior and Schwarzschild exterior
belong to compatible bulk branches and can be joined through a localized
source near the exterior root. The leading membrane treatment supplies the
corresponding matching conditions and a conditional restoring radial mode.
This does not yet constitute a complete time-dependent formation solution:
the resolved finite-width dynamics would require the full Einstein--Weyl
equations together with the evolution of the internal source and stress
profiles. The present result instead establishes a locally consistent
candidate endpoint with a regular interior and a timelike microscopic shell.

The non-Abelian fusion rules determine the state space carried by the shell.
At fixed constituent number, the constrained fusion-space dimension grows
as $\dim\mathcal H_N \sim d^N$,
so that $S_{\rm fus} = N\log d+O(1)$.
Identifying the constituent number with the horizon area then reproduces the
leading Bekenstein--Hawking entropy and fixes the relation between the area
per constituent and the quantum dimension. The discrete mass levels inherit
the fusion-space degeneracy, while the subleading dependence of the level
spacing produces a logarithmic correction whose coefficient is controlled by
\(\log d\). Ising and Fibonacci anyons provide explicit non-Abelian examples
with \(d=\sqrt2\) and \(d=\varphi\), respectively. Their values also satisfy
the heuristic local-mode bound \(d\geq e^{1/4}\) obtained in
Section~\ref{sec:thermodynamics}, although that bound depends on interpreting
\(4\log d\) literally as an effective number of local equipartition degrees
of freedom.

The same microscopic structure governs radiation. Emission is described by
transitions between discrete shell states rather than by entangled pair
creation across a causal horizon. The radiation need not therefore be exactly
Planckian: transition amplitudes and selection rules may retain information
about the fusion-channel state of the condensate. Information is not assigned
to an inaccessible singular region or to independent interior partners of
outgoing Hawking quanta, but remains encoded in the physical shell state and
can influence subsequent transitions.

The absorption analysis makes this radiation picture more restrictive than a
generic material-surface model. The fusion degeneracy fixes both the line
spacing and the reverse/absorption weight for a $k$-quantum transition. Identifying that rate ratio with reflectivity requires
the additional fluctuation--dissipation mapping used in quantum-horizon
response theory; it is not a microscopic transport derivation from the anyon
Hamiltonian. With that qualification, the same discrete response closes
the interior consistency condition: Hawking-order radiation emitted inward is
reabsorbed after only order-one crossings instead of accumulating into the
Tolman bath of a reflecting equilibrium cavity. The remaining microscopic
uncertainty concerns line widths, channel-dependent matrix elements, and the
partition of Hawking emission among exterior and interior modes.

The standard information paradox therefore does not arise in the present
model. Page curves and replica wormholes
\cite{AlmheiriEtAl2020Replica}, together with firewalls and black-hole
complementarity \cite{almheiri2013black}, address the tension produced by a
semiclassical description based on a causal horizon, interior partner states,
and pair creation across that horizon. These ingredients are absent here.
There is consequently no information hidden in an inaccessible interior that
must later be reconstructed or transferred at a Page time. Instead, the
information resides in the shell's fusion degrees of freedom and is
redistributed through transitions of the combined shell-radiation system. A
Page-like evolution of the radiation entropy may emerge from this dynamics,
but it would be a consequence of the microscopic evolution rather than a
mechanism required to restore information. The remaining task is to derive the
transition amplitudes and nonequilibrium fusion dynamics explicitly and
determine how the encoded fusion information appears in the outgoing
radiation.

The proposal is related to several earlier attempts to associate black-hole
entropy with microscopic degrees of freedom at or near the horizon. The
membrane paradigm \cite{thorne1986black}, conformal boundary-state counting
\cite{carlip1999black}, and the Chern--Simons description of isolated horizons
\cite{ashtekar2000isolated,pithis2015anyonic} all motivate some form of
horizon-localized dynamics. Anyonic statistics have also appeared in
connections between quantum geometry and horizon punctures
\cite{pithis2015anyonic,ghosh2014statistics}, while logarithmic entropy
corrections arise in loop quantum gravity \cite{kaul2000logarithmic}
and Euclidean quantum-gravity calculations \cite{sen2013logarithmic}. The present construction differs
in assigning the complete microscopic entropy and emission mechanism to a
physical, timelike non-Abelian condensate with a discrete fusion Hilbert space
at each mass level.

There is likewise a broad relation to holographic descriptions, since the
relevant information is encoded on a lower-dimensional region
\cite{hooft1993dimensional,susskind1995world}. The present mechanism does not,
however, require a bulk--boundary duality or AdS asymptotics. The shell is a
localized physical subsystem in an asymptotically flat geometry, and its
entropy follows directly from its fusion algebra. This distinguishes it from
AdS/CFT constructions \cite{maldacena1999large} while retaining the general
idea that gravitational information may admit a lower-dimensional
description.

The model also differs from graviton-condensate, gravastar, and fuzzball
proposals. Graviton \(N\)-portrait models describe a delocalized condensate of
soft gravitational quanta \cite{dvali2013black}, whereas the present
microscopic degrees of freedom are localized on a thin shell. Gravastars
replace the interior by a de~Sitter-like region whose vacuum energy contributes
substantially to the total mass and is joined to the exterior by a material
layer \cite{Mazur2004,Visser2004}. Here the regular interior is effectively
empty apart from its ambient curvature scale, while the shell and exterior
field carry the dominant mass. Fuzzball constructions replace the classical
black hole by an ensemble of horizon-scale string geometries
\cite{Mathur2005,Skenderis2008}; the present proposal instead retains a common
regular bulk geometry and assigns the state degeneracy to localized fusion
degrees of freedom. These comparisons concern the organization of the
microscopic state space rather than a claim that the different frameworks are
limits of one another.

The existence of a timelike shell also makes the usual observational
arguments for event horizons directly relevant. If an ultracompact object
possessed an ordinary material surface, accreted matter could form a
long-lived nuclear-burning layer and produce Type~I bursts
\cite{narayan2002lack}. Matter brought to rest on a conventional surface
would also release binding energy through an additional thermal component,
which is strongly constrained in systems such as Sgr~A*
\cite{broderick2009event}. The absence of these signatures is often
interpreted as evidence against material surfaces.

The condensate considered here is not an ordinary stellar surface. Infalling
energy couples to collective and fusion degrees of freedom in a highly
redshifted shell rather than being assumed to thermalize in a conventional
photosphere. Compatibility with the burst constraint requires that accreted
matter not accumulate as a persistent nuclear-burning layer. Compatibility
with surface-emission limits requires that deposited energy not be promptly
reradiated as an approximately thermal electromagnetic component with the
efficiency expected from ordinary matter.

The derived frequency dependence sharpens these observational requirements.
At frequencies well above the Hawking scale the shell is effectively as dark
as a classical horizon, whereas softer modes retain more reflectivity. The
latter can modify tidal absorption or generate delayed echoes, but the echo
amplitude is suppressed both by the shell response and by the exterior
curvature barrier. Electromagnetic branching fractions and the detailed fate
of accreted baryonic matter remain model dependent, so the absence of ordinary
surface emission continues to constrain the microscopic couplings rather than
the compactness alone.

The framework presently ties the microscopic and gravitational pictures to the
same physical shell: fusion counting supplies the entropy, the discrete
spectrum supplies the emission scale, equipartition reproduces the black-hole
temperature, the resulting thermal scale motivates a collective Hamiltonian
whose canonical entropy matches the same microscopic structure, and
Einstein--Weyl gravity supplies a regular near-horizon realization. The radial analysis is especially useful because it can be
compared directly with the corresponding timelike Israel shell rather than
with a different class of compact object. In general relativity the material
response required for support becomes singular as the shell is pushed toward
the Schwarzschild radius; in the Einstein--Weyl description the restoring
response remains finite there. Radial support is eventually lost at
a finite compactness that coincides with the Buchdahl radius. This coincidence
is physically suggestive: the stable membrane occupies the black-hole-like
side of the familiar stellar compactness scale, exactly where the proposed
endpoint of collapse is meant to live. The full derivations, including the
common Young--Laplace formulation, are collected in
Appendix~\ref{app:radial_stability}.

Several elements remain open. A microscopic completion must fix the precise
non-Abelian phase, the transition matrix elements and line widths that resolve
the coarse-grained absorption envelope, and the channel-dependent branching of
Hawking and electromagnetic emission. The correspondence between energy-level
degeneracy and the constrained fusion-space dimension may also receive
interaction-dependent corrections. On the gravitational side, the leading
membrane matching and monopole stability analysis must eventually be replaced
by a time-dependent finite-width solution of the full Einstein--Weyl system,
including perturbations of the internal source, stress profile, additional
normal support, and nonradial modes. These ingredients are needed
to turn the present effective response into a complete dynamical theory.

\section{Conclusion}

The model developed here replaces the would-be horizon by a physical,
topologically ordered timelike shell surrounding a regular, nearly flat
constant-curvature interior. Its constrained non-Abelian fusion space supplies the microscopic
state count, while the exterior remains Schwarzschild-like on macroscopic
scales. The construction therefore assigns entropy, information, and emission
to one localized many-body system rather than to a singular interior or an
abstract causal boundary.

The thermodynamic sector is internally linked. Fusion counting reproduces the
Bekenstein--Hawking area law, the associated discrete mass spectrum gives the
Hawking temperature with controlled subleading corrections, and the collective
Hamiltonian reproduces the leading and logarithmic entropy structure. The same
spectrum also constrains the dissipative response by fixing the allowed transition
spacings and reverse/absorption weights. When these are matched to the generic fluctuation--dissipation
response, resolved lines retain explicit anyonic structure, whereas strongly
overlapping lines approach the nearly universal Boltzmann absorption envelope.

The Einstein--Weyl construction supplies the complementary gravitational
realization. A finite-width anisotropic membrane can connect the regular
interior to the exterior near the Schwarzschild root, and its leading monopole
dynamics admit a finite positive-tension restoring response where the direct
Israel-shell benchmark becomes singular. This radial stability is conditional
and terminates at the Buchdahl compactness scale; nonradial and resolved
finite-width dynamics remain necessary before claiming complete stability.

The absorptive response also makes the regular interior self-consistent.
Hawking-type emission toward the interior need not be forbidden: it is
reabsorbed after only order-one crossings and therefore remains a dilute,
nonaccumulating component rather than forming the large Tolman equilibrium bath
of a reflecting cavity. The corresponding membrane-equivalent shear response approaches the
black-hole value only after the same response mapping is made; it is not a
microscopic viscosity derived from the anyon Hamiltonian. Residual
line reflectivity can feed suppressed gravitational-wave echoes
or modified tidal absorption. Thus black-hole-like darkness and a regular timelike interior are
compatible without imposing an exactly absorbing boundary by hand.

Because radiation is generated by transitions among physical shell states,
information remains in the shell--radiation system throughout the evolution.
The usual reconstruction of information from inaccessible interior partners is
therefore not the mechanism at work here. The unresolved microscopic task is
instead to determine how the fusion information enters the transition matrix
elements and hence the detailed spectrum, linewidths, branching fractions, and
late-time information flow.

A complete theory must still identify the underlying non-Abelian phase,
derive its dynamical couplings, and solve the time-dependent finite-width
Einstein--Weyl problem including nonradial perturbations. Within those limits,
the present construction ties together a microscopic area law, Hawking-scale
thermodynamics, conditional near-horizon support, and a calculable
coarse-grained absorption law in a single shell model, providing concrete
quantities by which the proposal can be distinguished from both an ordinary
material surface and a classical event horizon. In this sense, the model places black-hole microphysics in direct conceptual contact with topological quantum computation, with nonlocal fusion channels providing the same topologically protected information-bearing structure that underlies non-Abelian anyonic qubits.

\appendix

\section{Notation and entropy from the effective Hamiltonian}
\label{app:entropy}

\subsection{Symbols and radii}
\label{app:notation}

We collect here the notation used throughout the paper and in the derivations below, see Table \ref{table:1}.

\begin{table*}[]
\centering
\begingroup
\footnotesize
\setlength{\tabcolsep}{6pt}
\renewcommand{\arraystretch}{1.08}
\begin{tabular}{p{0.19\textwidth}p{0.75\textwidth}}
\hline
Symbol & Meaning \\
\hline

\multicolumn{2}{l}{\textit{Area spectrum, entropy, and thermodynamics}} \\[2pt]

\(A_0,\,N,\,n,\,N_{\rm dof}\) &
Area per constituent, constituent number, spectral level index, and
equipartition degree count. \\

\(d,\,u\) &
Quantum dimension and \(u=\log d\). \\

\(S_{\rm fus},\,S_{\rm therm},\,S_{\rm can}\) &
Fusion, thermodynamic, and canonical entropies. \\

\(M_n,\,\mu,\,\Delta E_n\) &
Discrete mass level, mass scale, and adjacent-level transition energy. \\

\(a,\,q,\,M_{\rm irr},\,E_{\rm therm}\) &
Dimensionless spin, charge-to-mass ratio, irreducible mass, and effective
thermal energy. \\

\(x_{\rm Kerr},\,x_{\rm RN}\) &
Irreducible mass normalized by its extremal value in the Kerr and
Reissner--Nordstr\"om families. \\[4pt]

\multicolumn{2}{l}{\textit{Collective Hamiltonian variables}} \\[2pt]

\(\mathbf{x}_i,\,\bar{\mathbf{x}},\,\mathbf{y}_i\) &
Collective coordinate, mean collective coordinate, and relative coordinate. \\

\(m,\,D,\,\tilde r,\,r_0\) &
Local component number, relative-space dimension, relative RMS scale, and
microscopic patch radius. \\

\(z,\,z_0,\,\widehat{\kappa},\,\widehat{\lambda}\) &
Dimensionless radial variable, preferred radial value, and Planck-rescaled
relative and radial stiffnesses. \\

\(\kappa,\,\lambda,\,\eta,\,p\) &
Relative-mode stiffness, radial-constraint stiffness, center-fixing
stiffness, and large-\(N\) center-fixing exponent. \\

\(\alpha_1\) &
Leading extensive coefficient of the canonical entropy. \\[4pt]

\multicolumn{2}{l}{\textit{Conformal-gravity geometry and membrane location}} \\[2pt]

\(M_P,\,\alpha_g,\,W^\mu{}_\nu\) &
Reduced Planck mass, conformal coupling, and Bach tensor. \\

\(B_{\rm in}(r),\,B_{\rm out}(r)\) &
Regular interior and exterior Mannheim--Kazanas bulk functions. \\

\(\rho_{\rm in},\,p_{\rm in},\,\Lambda_{\rm in},\,\Delta\Lambda_{\rm in}\) &
Signed interior vacuum-energy density, isotropic interior pressure, geometric
interior cosmological constant, and its effective curvature correction. \\

\(r_c,\,r_s,\,B_c,\,\delta\) &
Membrane radius, exterior root, exterior lapse at the membrane, and
near-horizon parameter, with \(B_c=\delta\). \\

\(h,\,\ell_c,\,\kappa_s\) &
Proper membrane thickness, proper membrane--root gap, and exterior
surface-gravity scale. \\[4pt]

\multicolumn{2}{l}{\textit{Membrane source and radial response}} \\[2pt]

\(\Sigma,\,\Sigma_c\) &
Proper projected Bach source and its equilibrium value. \\

\(\sigma_{\rm CG},\,p_s,\,a\) &
Physical surface energy density, positive tangential tension, and integrated
radial-anisotropy parameter. \\

\(\rho,\,p_r,\,p_t,\,T\) &
Bulk energy density, radial and tangential pressures, and stress-tensor trace. \\

\(\overline K^\tau{}_\tau,\,\overline H\) &
Averaged temporal and angular extrinsic curvatures entering the normal-force
balance. \\

\(\eta_{\rm CG},\,\eta_{\rm CG}^{\rm crit}\) &
Positive-tension constitutive response and its critical value for radial
stability. \\

\(r(\tau),\,\omega_{\rm CG}\) &
Dynamical membrane radius and proper-time radial frequency. \\

\(\eta_{\rm GR},\,\eta_{\rm GR}^{\rm crit}\) &
Israel-shell constitutive response and its critical value for radial
stability. \\[4pt]

\multicolumn{2}{l}{\textit{Absorption and echoes}} \\[2pt]

\(g_\ast,\,\alpha_{\rm leak}\) &
Effective massless-field count and inward Hawking-leakage coefficient. \\

\(\mathcal R_{\rm sh},\,\mathcal A_{\rm sh}\) &
Membrane reflection amplitude and absorptivity. \\

\(\eta_{\rm visc}^{\rm eff},\,\eta_{\rm BH}\) &
Membrane-equivalent shear viscosity inferred from reflectivity, and the
classical membrane-paradigm value. \\

\(\tau_{\rm res}^{\infty},\,U_{\rm int}^{\infty}\) &
Exterior-time residence scale and redshifted stored interior energy. \\

\(\Delta\omega,\,\mathcal R_{\rm bar},\,\Delta t_{\rm echo}\) &
Microscopic line spacing, exterior-barrier reflection amplitude, and echo delay. \\

\hline
\end{tabular}
\endgroup
\caption{Notation used throughout the paper. The membrane denotes the
finite-width anisotropic transition layer, while the shell denotes its
effective thin-wall or low-energy description.}
\label{table:1}
\end{table*}

\subsection{Entropy from the Effective Hamiltonian}
\label{app:entropy_hamiltonian}

In this appendix we derive the large-\(N\) entropy associated with the
effective Hamiltonian description of the horizon-shell degrees of freedom used
in the main text. The derivation assumes the large-\(N\) limit, a microscopic
fluctuation scale \(\tilde r^{\,2}\simeq r_0^2\), Gaussian fixing of the
average collective coordinate, and a finite radial constraint that localizes
the relative collective modes around \(r_0\). Throughout, \(Z(\beta)\) is a
configurational, rather than phase-space, partition function. The
configurational measure is normalized relative to the microscopic length
\(\ell_P\), so that the partition function and the arguments of all logarithms
are dimensionless.

We consider \(N\) collective constituents with \(m\) components
\(\mathbf{x}_i\in\mathbb{R}^m\). Define the average coordinate and relative
mean-square scale by
\begin{equation}
    \bar{\mathbf{x}}
    =
    \frac{1}{N}\sum_{i=1}^{N}\mathbf{x}_i,
    \qquad
    \tilde r^{\,2}
    =
    \frac{1}{N}
    \sum_{i=1}^{N}
    \left\|
        \mathbf{x}_i-\bar{\mathbf{x}}
    \right\|^2 .
\end{equation}
The effective Hamiltonian is taken to be
\begin{equation}
\begin{aligned}
    H\!\left(\{\mathbf{x}_i\}\right)
    ={}&
    \frac{\kappa}{2}
    \sum_{i=1}^{N}
    \left\|
        \mathbf{x}_i-\bar{\mathbf{x}}
    \right\|^2
    +
    \frac{\lambda N}{4}
    \left(
        \tilde r^{\,2}-r_0^2
    \right)^2
    \\
    &+
    \frac{\eta N^{p+1}}{2}
    \left\|
        \bar{\mathbf{x}}
    \right\|^2 .
\end{aligned}
\label{eq:app_effective_hamiltonian_corrected}
\end{equation}
Here \(\kappa>0\) is the stiffness of the relative quadratic modes,
\(\lambda>0\) controls radial fluctuations around \(r_0\), and \(\eta>0\)
fixes the average collective coordinate. The factor \(N^{p+1}\) is chosen
because \(\bar{\mathbf{x}}\) is the average coordinate. The collective
translation direction has configuration-space line element
\(N\|d\bar{\mathbf{x}}\|^2\), and this factor must be retained when
logarithmic powers of \(N\) are tracked. With this convention, \(p\) controls
the large-\(N\) contribution of the center-fixing term to the entropy.

The configurational partition function is
\begin{equation}
    Z(\beta)
    =
    \int_{\mathbb{R}^{mN}}
    \prod_{i=1}^{N}
    \frac{d^m x_i}{\ell_P^m}\,
    e^{-\beta H}.
\label{eq:app_dimensionless_partition}
\end{equation}
Introduce relative coordinates
\begin{equation}
    \mathbf{y}_i
    =
    \mathbf{x}_i-\bar{\mathbf{x}},
    \qquad
    \sum_{i=1}^{N}\mathbf{y}_i=0 .
\end{equation}
This separates the \(mN\) coordinates into the \(m\) components of
\(\bar{\mathbf{x}}\) and
\begin{equation}
    D
    :=
    m(N-1)
\end{equation}
relative coordinates spanning the hyperplane
\(\sum_i\mathbf{y}_i=0\). The configuration-space measure becomes
\begin{equation}
    \prod_{i=1}^{N}
    \frac{d^m x_i}{\ell_P^m}
    =
    N^{m/2}
    \frac{d^m\bar x}{\ell_P^m}
    \frac{d^D y}{\ell_P^D},
\label{eq:app_measure_barx}
\end{equation}
where \(d^D y\) denotes the induced Euclidean measure on the relative
subspace. The factor \(N^{m/2}\) is the Jacobian associated with using the
average coordinate rather than the unit-normalized collective coordinate.

In these variables,
\begin{equation}
    \tilde r^{\,2}
    =
    \frac{1}{N}
    \sum_{i=1}^{N}
    \|\mathbf{y}_i\|^2 .
\end{equation}
The Hamiltonian separates into center and relative parts,
\begin{equation}
    H
    =
    H_{\rm cm}
    +
    H_{\rm rel},
\end{equation}
with
\begin{equation}
    H_{\rm cm}
    =
    \frac{\eta N^{p+1}}{2}
    \left\|
        \bar{\mathbf{x}}
    \right\|^2,
\end{equation}
and
\begin{equation}
    H_{\rm rel}
    =
    \frac{\kappa}{2}
    \sum_{i=1}^{N}\|\mathbf{y}_i\|^2
    +
    \frac{\lambda N}{4}
    \left(
        \tilde r^{\,2}-r_0^2
    \right)^2 .
\end{equation}
The partition function therefore factorizes as
\begin{equation}
    Z(\beta)
    =
    Z_{\rm cm}(\beta)
    Z_{\rm rel}(\beta),
\end{equation}
up to \(N\)-independent normalization constants.

The center Gaussian, including the Jacobian in
Eq.~\eqref{eq:app_measure_barx}, gives
\begin{equation}
\begin{aligned}
    Z_{\rm cm}(\beta)
    &=
    N^{m/2}
    \int_{\mathbb{R}^{m}}
    \frac{d^m\bar x}{\ell_P^m}
    \exp\!\left[
        -\beta
        \frac{\eta N^{p+1}}{2}
        \left\|
            \bar{\mathbf{x}}
        \right\|^2
    \right]
    \\
    &=
    N^{m/2}
    \left(
        \frac{2\pi}
        {\beta\eta N^{p+1}\ell_P^2}
    \right)^{m/2}.
\end{aligned}
\end{equation}
Hence
\begin{equation}
    \log Z_{\rm cm}(\beta)
    =
    -\frac{pm}{2}\log N
    +
    O(1),
\label{eq:app_cm_log}
\end{equation}
where \(O(1)\) includes all \(\beta\)-dependent terms and \(N\)-independent
normalization constants.

For the relative sector, define the relative configuration-space radius
\begin{equation}
    s^2
    :=
    \sum_{i=1}^{N}\|\mathbf{y}_i\|^2
    =
    N\tilde r^{\,2},
\label{eq:app_relative_radius}
\end{equation}
and the preferred radial value
\begin{equation}
    s_N
    :=
    \sqrt{N}\,r_0 .
\label{eq:app_relative_shell_radius}
\end{equation}
It is convenient to introduce the dimensionless variables
\begin{equation}
    z
    :=
    \frac{\tilde r^{\,2}}{\ell_P^2}
    =
    \frac{s^2}{N\ell_P^2},
    \qquad
    z_0
    :=
    \frac{r_0^2}{\ell_P^2},
\label{eq:app_dimensionless_radial_variables}
\end{equation}
together with the Planck-rescaled stiffnesses
\begin{equation}
    \widehat{\kappa}
    :=
    \kappa\ell_P^2,
    \qquad
    \widehat{\lambda}
    :=
    \lambda\ell_P^4.
\label{eq:app_rescaled_stiffnesses}
\end{equation}
The relative Hamiltonian can then be written as
\begin{equation}
    H_{\rm rel}
    =
    Nh(z),
\end{equation}
where
\begin{equation}
    h(z)
    =
    \frac{\widehat{\kappa}}{2}z
    +
    \frac{\widehat{\lambda}}{4}
    (z-z_0)^2.
\label{eq:app_radial_hamiltonian}
\end{equation}

The radial form of the relative measure is
\begin{equation}
    \frac{d^D y}{\ell_P^D}
    =
    \Omega_D
    \left(
        \frac{s}{\ell_P}
    \right)^{D-1}
    \frac{ds}{\ell_P}
    =
    \frac{\Omega_D}{2}
    N^{D/2}
    z^{D/2-1}dz,
\label{eq:app_radial_measure}
\end{equation}
where
\begin{equation}
    \Omega_D
    =
    \frac{2\pi^{D/2}}{\Gamma(D/2)}
\end{equation}
is the area of the unit \((D-1)\)-sphere. The relative partition function is
therefore
\begin{equation}
    Z_{\rm rel}(\beta)
    =
    \frac{\Omega_D}{2}
    N^{D/2}
    \int_0^\infty
    dz\,
    z^{D/2-1}
    e^{-\beta Nh(z)}.
\label{eq:app_Zrel_surface}
\end{equation}

Since \(D=m(N-1)\), the leading large-\(N\) radial integral has the saddle
form
\begin{equation}
    Z_{\rm rel}
    \sim
    \frac{\Omega_D}{2}
    N^{D/2}
    \int_0^\infty dz\,
    e^{N\Phi(z)},
\end{equation}
where
\begin{equation}
    \Phi(z)
    =
    \frac{m}{2}\log z
    -
    \beta
    \left[
        \frac{\widehat{\kappa}}{2}z
        +
        \frac{\widehat{\lambda}}{4}
        (z-z_0)^2
    \right].
\label{eq:app_soft_saddle_function}
\end{equation}
The saddle \(z=z_\ast\) satisfies
\begin{equation}
    \Phi'(z_\ast)=0,
\end{equation}
or
\begin{equation}
    \boxed{
    \beta
    \left[
        \widehat{\kappa}
        +
        \widehat{\lambda}(z_\ast-z_0)
    \right]
    =
    \frac{m}{z_\ast}.
    }
\label{eq:app_soft_saddle_equation}
\end{equation}
Its physical solution is
\begin{equation}
    z_\ast
    =
    \frac{
        \widehat{\lambda}z_0-\widehat{\kappa}
        +
        \sqrt{
            \left(
                \widehat{\kappa}
                -
                \widehat{\lambda}z_0
            \right)^2
            +
            4\widehat{\lambda}m/\beta
        }
    }{
        2\widehat{\lambda}
    },
\label{eq:app_soft_saddle_solution}
\end{equation}
where the positive branch is selected because \(z_\ast\geq0\). The saddle is
locally stable since
\begin{equation}
    \Phi''(z_\ast)
    =
    -\frac{m}{2z_\ast^2}
    -
    \frac{\beta\widehat{\lambda}}{2}
    <
    0.
\end{equation}

For large \(D\), Stirling's approximation gives
\begin{equation}
    \log\Omega_D
    =
    \frac{D}{2}
    \left[
        1+
        \log\!\left(
            \frac{2\pi}{D}
        \right)
    \right]
    +
    \frac{1}{2}\log D
    +
    O(1).
\end{equation}
Evaluating the radial integral around \(z_\ast\) then gives
\begin{equation}
\begin{aligned}
    \log Z_{\rm rel}
    ={}&
    N
    \left\{
        \frac{m}{2}
        \left[
            1+
            \log\!\left(
                \frac{2\pi z_\ast}{m}
            \right)
        \right]
        -
        \beta h(z_\ast)
    \right\}
    +
    O(1).
\end{aligned}
\label{eq:app_logZrel_final}
\end{equation}
The Gaussian radial integral contributes
\(-\tfrac12\log N\), while the subleading
\(\tfrac12\log D\) term in \(\log\Omega_D\) contributes
\(+\tfrac12\log N\). Since \(D=m(N-1)\), these terms cancel, and the relative
sector produces no independent logarithmic contribution.

Define
\begin{equation}
    \alpha_1(z_\ast)
    :=
    \frac{m}{2}
    \left[
        1+
        \log\!\left(
            \frac{2\pi z_\ast}{m}
        \right)
    \right].
\label{eq:app_alpha1}
\end{equation}
Combining Eqs.~\eqref{eq:app_cm_log} and
\eqref{eq:app_logZrel_final}, the full configurational partition function
obeys
\begin{equation}
    \log Z(\beta)
    =
    \alpha_1(z_\ast)N
    -
    \frac{pm}{2}\log N
    -
    \beta Nh(z_\ast)
    +
    O(1).
\label{eq:app_logZ_total}
\end{equation}

In units with \(k_{\rm B}=1\), the canonical entropy is
\begin{equation}
    S_{\rm can}
    =
    \beta\langle H\rangle+\log Z .
\end{equation}
At leading order,
\begin{equation}
    \langle H_{\rm rel}\rangle
    =
    Nh(z_\ast)
    +
    O(1),
\end{equation}
while the center Gaussian contributes only an \(O(1)\) term,
\begin{equation}
    \langle H_{\rm cm}\rangle
    =
    \frac{m}{2\beta}.
\end{equation}
The saddle equation ensures that the implicit \(\beta\)-dependence of
\(z_\ast\) does not contribute at leading order. The extensive relative-energy
term therefore cancels the explicit \(-\beta Nh(z_\ast)\) contribution in
\(\log Z\), giving
\begin{equation}
    \boxed{
    S_{\rm can}(N)
    =
    \alpha_1(z_\ast)N
    -
    \frac{pm}{2}\log N
    +
    O(1).
    }
\label{eq:app_entropy_general}
\end{equation}

The microscopic patch area is
\begin{equation}
    A_0
    =
    4\pi r_0^2
    =
    4\ell_P^2\log d .
\end{equation}
Define
\begin{equation}
    u
    :=
    \log d,
    \qquad
    z_0
    =
    \frac{r_0^2}{\ell_P^2}
    =
    \frac{u}{\pi}.
\end{equation}
The identification of \(r_0\) as the equilibrium microscopic patch scale
corresponds to
\begin{equation}
    z_\ast=z_0 .
\label{eq:app_equilibrium_radius}
\end{equation}
Equation~\eqref{eq:app_alpha1} then becomes
\begin{equation}
    \alpha_1(u)
    =
    \frac{m}{2}
    \left(
        1+\log\frac{2u}{m}
    \right).
\end{equation}
Matching the leading extensive entropy density to the fusion entropy: $\alpha_1=u$, gives
\begin{equation}
    u
    =
    \frac{m}{2}
    \left(
        1+\log\frac{2u}{m}
    \right).
\end{equation}
The positive solution is
\begin{equation}
    \boxed{
    u
    =
    \log d
    =
    \frac{m}{2},
    \qquad
    d
    =
    (\sqrt e)^{\,m}.
    }
\label{eq:app_d_relation}
\end{equation}
Consequently,
\begin{equation}
    z_0
    =
    \frac{m}{2\pi},
    \qquad
    r_0^2
    =
    \frac{m}{2\pi}\ell_P^2 .
\label{eq:app_matched_patch_radius}
\end{equation}

The relative-mode stiffness is identified with the exterior surface-gravity
scale through
\begin{equation}
    \widehat{\kappa}
    =
    \kappa\ell_P^2
    =
    \kappa_s.
\label{eq:app_surface_gravity_stiffness}
\end{equation}
At the Hawking temperature,
\begin{equation}
    T_H
    =
    \frac{\kappa_s}{2\pi},
    \qquad
    \beta_H
    =
    \frac{2\pi}{\kappa_s},
\label{eq:app_Hawking_surface_gravity}
\end{equation}
one has
\begin{equation}
    \beta_H\widehat{\kappa}z_0
    =
    \frac{2\pi}{\kappa_s}
    \kappa_s
    \frac{m}{2\pi}
    =
    m.
\end{equation}
The radial saddle equation
\eqref{eq:app_soft_saddle_equation} therefore becomes
\begin{equation}
    z
    \left[
        \widehat{\kappa}
        +
        \widehat{\lambda}(z-z_0)
    \right]
    =
    \widehat{\kappa}z_0,
\end{equation}
which factors as
\begin{equation}
    (z-z_0)
    \left(
        \widehat{\lambda}z
        +
        \widehat{\kappa}
    \right)
    =
    0.
\label{eq:app_exact_soft_factorization}
\end{equation}
For
\(\widehat{\kappa}>0\), \(\widehat{\lambda}>0\), and \(z\geq0\), the second
root \(z=-\widehat{\kappa}/\widehat{\lambda}\) is excluded. The physical
saddle is therefore $z_\ast=z_0$ for finite positive \(\widehat{\lambda}\).

The thermodynamic entropy derived from the discrete area spectrum has the form
\begin{equation}
    S_{\rm therm}(N)
    =
    N\log d
    -
    \frac{\log d}{4}\log N
    +
    O(1).
\end{equation}
Comparing this with Eq.~\eqref{eq:app_entropy_general}, and using
\(\log d=m/2\), gives
\begin{equation}
    \frac{pm}{2}
    =
    \frac{\log d}{4}
    =
    \frac{m}{8}.
\end{equation}
Thus $p = \cfrac{1}{4}$. With this choice,
\begin{equation}
    \boxed{
    \begin{aligned}
    S_{\rm can}(N)
    & =
    \frac{m}{2}N
    -
    \frac{m}{8}\log N
    +
    O(1)
    \\
    & =
    N\log d
    -
    \frac{\log d}{4}\log N
    +
    O(1)
    \end{aligned}
    }.
\label{eq:app_entropy_final}
\end{equation}
The relation \(d=(\sqrt e)^{\,m}\) is the semiclassical matching condition of
the effective collective model. The relative shell modes determine the leading
extensive entropy, while the large-\(N\) scaling of the Gaussian
center-coordinate fixing determines the logarithmic term.

\section{Detailed radial stability derivations}
\label{app:radial_stability}

This appendix gives the radial stability calculations underlying
Section~\ref{subsec:membrane_response}. The purpose is twofold. First, the
Minkowski-interior/Schwarzschild-exterior shell is treated entirely within
general relativity, both by the standard Israel junction-potential method and
by the generalized Young--Laplace normal-balance equation. The two routes give
the same critical equation-of-state response. Second, the same Young--Laplace
construction is applied to the Einstein--Weyl membrane, making explicit where
the projected Bach source and the positive tangential tension modify the
near-horizon response.

\subsection{Israel shell: exact-potential and Young--Laplace derivations}
\label{app:israel_stability}

Consider a timelike spherical shell of areal radius \(r(\tau)\), with a
Minkowski interior and Schwarzschild exterior,
\begin{equation}
    B_-(r)=1,
    \qquad
    B_+(r)=f(r)=1-\frac{r_s}{r}.
\label{eq:app_israel_metrics}
\end{equation}
Define
\begin{equation}
    \beta_-:=\sqrt{\dot r^{\,2}+1},
    \qquad
    \beta_+:=\sqrt{\dot r^{\,2}+f(r)}.
\label{eq:app_israel_betas}
\end{equation}
The physical surface stress tensor is
\begin{equation}
    S_{\hat a\hat b}
    =
    \operatorname{diag}
    \left(\sigma_{\rm GR},p_t,p_t\right),
\end{equation}
where \(p_t>0\) denotes tangential pressure. Equivalently, in the
positive-tension convention of the main text, \(p_s=-p_t<0\). No bulk normal
pressure is present on either side,
\begin{equation}
    p_{r,\rm in}=p_{r,\rm out}=0.
\label{eq:app_israel_bulk_pressure}
\end{equation}

\subsubsection{Short route: exact Israel potential}

The angular Israel junction condition can be written as
\begin{equation}
    \beta_- - \beta_+
    =
    4\pi G r\,\sigma_{\rm GR}
    \equiv
    \zeta(r).
\label{eq:app_israel_angular_jump}
\end{equation}
Solving Eq.~\eqref{eq:app_israel_angular_jump} for the radial velocity gives
\begin{equation}
    \dot r^{\,2}+V(r)=0,
\end{equation}
with
\begin{equation}
    V(r)
    =
    \frac{1+f}{2}
    -\frac{\zeta^2}{4}
    -\frac{(1-f)^2}{4\zeta^2}.
\label{eq:app_israel_potential}
\end{equation}
A static configuration at \(r=r_c\) satisfies
\(V_c=V_c'=0\). Introduce
\begin{equation}
    x:=\sqrt{\delta},
    \qquad
    \delta=f(r_c)=1-\frac{r_s}{r_c},
    \qquad 0<x<1.
\label{eq:app_israel_x}
\end{equation}
The static junction conditions then give
\begin{equation}
    \zeta_c=1-x,
\end{equation}
\begin{equation}
    \sigma_{{\rm GR},c}
    =
    \frac{1-x}{4\pi G r_c},
    \qquad
    p_{t,c}
    =
    \frac{(1-x)^2}{16\pi G r_c x}.
\label{eq:app_israel_static_stresses}
\end{equation}
The intrinsic surface conservation law is
\begin{equation}
    \frac{d}{d\tau}
    \left(4\pi r^2\sigma_{\rm GR}\right)
    +
    p_t\frac{d}{d\tau}(4\pi r^2)
    =0,
\end{equation}
which, wherever \(r\) can be used as the local radial variable, becomes
\begin{equation}
    \sigma_{\rm GR}'
    =
    -\frac{2}{r}
    \left(\sigma_{\rm GR}+p_t\right).
\label{eq:app_israel_conservation}
\end{equation}
With
\begin{equation}
    \eta_{\rm GR}
    :=
    \left.\frac{dp_t}{d\sigma_{\rm GR}}\right|_{r_c},
\end{equation}
one has \(p_t'=\eta_{\rm GR}\sigma_{\rm GR}'\). From
\(\zeta=4\pi G r\sigma_{\rm GR}\), Eqs.~\eqref{eq:app_israel_static_stresses}
--\eqref{eq:app_israel_conservation} give
\begin{equation}
    \zeta_c'
    =
    -\frac{1-x^2}{2r_c x},
\end{equation}
\begin{equation}
    \zeta_c''
    =
    \frac{(1-x)(1+3x)}{2r_c^2x}
    \left(1+2\eta_{\rm GR}\right).
\label{eq:app_israel_zeta_derivatives}
\end{equation}
Substituting these derivatives into the second derivative of
Eq.~\eqref{eq:app_israel_potential} yields
\begin{equation}
    V_c''
    =
    \frac{1}{2r_c^2x^2}
    \left[
        4\eta_{\rm GR}x^2(1+3x)
        +3x^3-x^2-x-1
    \right].
\label{eq:app_israel_Vpp}
\end{equation}
The static shell is radially stable when \(V_c''>0\). Since the prefactor in
Eq.~\eqref{eq:app_israel_Vpp} is positive,
\begin{equation}
    \boxed{
    \eta_{\rm GR}
    >
    \eta_{\rm GR}^{\rm crit}(\delta)
    =
    \frac{
        1+\sqrt\delta+\delta-3\delta^{3/2}
    }{
        4\delta(1+3\sqrt\delta)
    }
    }.
\label{eq:app_israel_threshold}
\end{equation}
Thus
\begin{equation}
    \eta_{\rm GR}^{\rm crit}
    \sim
    \frac{1}{4\delta}
    \longrightarrow +\infty,
    \qquad \delta\to0^+.
\label{eq:app_israel_near_horizon}
\end{equation}
This is the result quoted in Eq.~\eqref{eq:etaGR_comparison}.

\subsubsection{Long route: generalized Young--Laplace balance}

The same result follows directly from the relativistic Young--Laplace equation,
which is the form used for the Einstein--Weyl membrane. For the Israel shell,
Eq.~\eqref{eq:general_normal_balance} has
\(p_{r,\rm in}-p_{r,\rm out}=0\) and \(p_s=-p_t\), and therefore
\begin{equation}
    0
    =
    \sigma_{\rm GR}\,\overline K^\tau{}_{\tau}
    -2p_t\overline H.
\label{eq:app_israel_YL}
\end{equation}
Using Eq.~\eqref{eq:app_israel_betas},
\begin{equation}
\begin{aligned}
    0={}&
    \frac{\sigma_{\rm GR}}{2}
    \left[
        \frac{\ddot r+f'/2}{\beta_+}
        +
        \frac{\ddot r}{\beta_-}
    \right]
    -
    \frac{p_t}{r}
    \left(\beta_++\beta_-\right).
\end{aligned}
\label{eq:app_israel_YL_dynamic}
\end{equation}
Velocity corrections begin at order \(\dot r^{\,2}\), so for
\(r=r_c+\xi(\tau)\) they do not enter the linearized equation. Define
\begin{equation}
    g(r):=\frac{f'(r)}{4\sqrt{f(r)}},
    \qquad
    h(r):=\frac{1+\sqrt{f(r)}}{r}.
\label{eq:app_israel_gh}
\end{equation}
The static part of Eq.~\eqref{eq:app_israel_YL_dynamic} is
\begin{equation}
    \mathcal F_{\rm GR}(r)
    :=
    \sigma_{\rm GR}(r)g(r)-p_t(r)h(r),
    \qquad
    \mathcal F_{{\rm GR},c}=0.
\label{eq:app_israel_force}
\end{equation}
The linearized radial equation is
\begin{equation}
    \boxed{
    \frac{\sigma_{{\rm GR},c}}{2}
    \left(1+\frac{1}{x}\right)
    \ddot\xi
    +
    \mathcal F_{{\rm GR},c}'\,\xi
    =0.
    }
\label{eq:app_israel_linearized_YL}
\end{equation}
The inertial coefficient is positive for \(\sigma_{{\rm GR},c}>0\), so
stability is equivalent to \(\mathcal F_{{\rm GR},c}'>0\). Differentiating
Eq.~\eqref{eq:app_israel_force} and using
Eq.~\eqref{eq:app_israel_conservation} gives
\begin{equation}
    \mathcal F_{{\rm GR},c}'
    =
    \sigma_{{\rm GR},c}'
    \left(g_c-\eta_{\rm GR}h_c\right)
    +
    \sigma_{{\rm GR},c}g_c'
    -
    p_{t,c}h_c'.
\label{eq:app_israel_force_prime}
\end{equation}
At \(f(r_c)=x^2\),
\begin{equation}
    g_c=\frac{1-x^2}{4r_cx},
    \qquad
    h_c=\frac{1+x}{r_c},
\end{equation}
\begin{equation}
    g_c'
    =
    -\frac{(1-x^2)(1+3x^2)}{8r_c^2x^3},
    \qquad
    h_c'
    =
    -\frac{(1+x)(3x-1)}{2r_c^2x}.
\label{eq:app_israel_gh_derivatives}
\end{equation}
Substitution of Eqs.~\eqref{eq:app_israel_static_stresses},
\eqref{eq:app_israel_conservation}, and
\eqref{eq:app_israel_gh_derivatives} into
Eq.~\eqref{eq:app_israel_force_prime} gives
\begin{equation}
    \boxed{
\begin{aligned}
    \mathcal F_{{\rm GR},c}'
    ={}&
    \frac{1-x^2}{32\pi G r_c^3x^3}
    \Bigl[
        4\eta_{\rm GR}x^2(1+3x)
        \\
        &\hspace{4.2em}
        +3x^3-x^2-x-1
    \Bigr].
\end{aligned}
    }
\label{eq:app_israel_force_prime_explicit}
\end{equation}
Every prefactor outside the square bracket is positive for \(0<x<1\).
Therefore \(\mathcal F_{{\rm GR},c}'>0\) reproduces exactly
Eq.~\eqref{eq:app_israel_threshold}. The exact Israel-potential method and the
generalized Young--Laplace method are thus equivalent for the radial mode; the
second form is useful because it can be carried over directly to the
Einstein--Weyl membrane.

\subsection{Einstein--Weyl membrane from generalized Young--Laplace balance}
\label{app:cg_stability}

For the Einstein--Weyl membrane the physical surface tensor is
Eq.~\eqref{eq:physical_surface_tensor},
\begin{equation}
    S_{\hat a\hat b}
    =
    \operatorname{diag}
    \left(\sigma_{\rm CG},-p_s,-p_s\right),
    \qquad p_s>0,
\end{equation}
and the normal-force equation is
\begin{equation}
    p_{r,\rm in}-p_{r,\rm out}
    =
    \sigma_{\rm CG}\,\overline K^\tau{}_{\tau}
    +
    2p_s\overline H.
\label{eq:app_cg_YL}
\end{equation}
The radial-pressure difference is generally nonzero at equilibrium. Only its
first radial variation is neglected for the constant-curvature interior and
vacuum exterior used in the leading stability calculation,
\begin{equation}
    \left.
    \frac{d}{dr}\left(p_{r,\rm in}-p_{r,\rm out}\right)
    \right|_{r_c}=0.
\label{eq:app_cg_radial_pressure_derivative}
\end{equation}
Thus the positive tension is balanced by the background bulk support rather
than by setting the radial-pressure difference itself to zero.

To the same weak-interior-curvature order used in the main text, we set
\(B_{\rm in}(r_c)\simeq1\) and neglect \(B_{\rm in}\prime(r_c)\) relative to
the Schwarzschild-side gradient. With \(B_{\rm out}=f=1-r_s/r\),
Eq.~\eqref{eq:app_cg_YL} becomes
\begin{equation}
\begin{aligned}
    p_{r,\rm in}-p_{r,\rm out}={}&
    \frac{\sigma_{\rm CG}}{2}
    \left[
        \frac{\ddot r+f'/2}
             {\sqrt{\dot r^{\,2}+f}}
        +
        \frac{\ddot r}
             {\sqrt{\dot r^{\,2}+1}}
    \right]
    \\
    &+
    \frac{p_s}{r}
    \left[
        \sqrt{\dot r^{\,2}+f}
        +
        \sqrt{\dot r^{\,2}+1}
    \right].
\end{aligned}
\label{eq:app_cg_YL_dynamic}
\end{equation}
Using the same auxiliary functions as in the main text,
\begin{equation}
    g(r)=\frac{f'}{4\sqrt f},
    \qquad
    h(r)=\frac{1+\sqrt f}{r},
\end{equation}
define the static force function
\begin{equation}
    \mathcal F_{\rm CG}(r)
    :=
    \sigma_{\rm CG}(r)g(r)
    +
    p_s(r)h(r)
    -
    \left[p_{r,\rm in}(r)-p_{r,\rm out}(r)\right].
\label{eq:app_cg_force}
\end{equation}
The background equation is \(\mathcal F_{{\rm CG},c}=0\). Writing
\(r=r_c+\xi(\tau)\), the terms containing \(\dot r\) again begin at quadratic
order. The coefficient of \(\ddot\xi\) is
\begin{equation}
    \mathcal A_c
    =
    \frac{\sigma_{{\rm CG},c}}{2}
    \left(1+\delta^{-1/2}\right)>0,
\label{eq:app_cg_inertial_coefficient}
\end{equation}
so
\begin{equation}
    \mathcal A_c\ddot\xi
    +
    \mathcal F_{{\rm CG},c}'\xi
    =0,
    \qquad
    \omega_{\rm CG}^2
    =
    \frac{\mathcal F_{{\rm CG},c}'}{\mathcal A_c}.
\label{eq:app_cg_linearized}
\end{equation}
Hence the leading radial mode is stable precisely when
\begin{equation}
    \mathcal F_{{\rm CG},c}'>0.
\label{eq:app_cg_stability_sign}
\end{equation}

The radial dependence of the projected source follows from the full
Einstein--Weyl equation. In either one-function bulk region,
Eq.~\eqref{eq:masterIII} gives the leading proper source moment
\begin{equation}
    \Sigma(r)
    =
    \frac{4\alpha_g}{3}
    \sqrt{f(r)}\,f^{(3)}(r).
\end{equation}
For Schwarzschild,
\begin{equation}
    \Sigma(r)
    =
    \frac{8\alpha_g r_s}{r^4}\sqrt{f(r)}.
\label{eq:app_cg_Sigma}
\end{equation}
Taking one radial derivative before imposing \(r=r_c\) gives
\begin{equation}
    \frac{\Sigma'}{\Sigma}
    =
    \frac{f'}{2f}-\frac{4}{r},
\end{equation}
and therefore
\begin{equation}
    \Sigma_c'
    =
    \frac{\Sigma_c}{2r_c\delta}
    \left(1-9\delta\right)
    .
\label{eq:app_cg_Sigmaprime}
\end{equation}
The factor \(1-9\delta\) is the origin of the finite-radius endpoint discussed
below.

The physical surface energy density differs from this projected source. The
closure used in the main text is
\begin{equation}
    \sigma_{\rm CG}
    =
    \Sigma+a p_s,
\label{eq:app_cg_closure}
\end{equation}
with fixed \(a\) under the leading perturbation, and
\begin{equation}
    \eta_{\rm CG}
    :=
    \left.\frac{dp_s}{d\sigma_{\rm CG}}\right|_{r_c}.
\end{equation}
Differentiating Eq.~\eqref{eq:app_cg_closure} gives
\begin{equation}
    \sigma_{\rm CG}'
    =
    \Sigma'+a p_s'
    =
    \Sigma'+a\eta_{\rm CG}\sigma_{\rm CG}',
\end{equation}
and hence
\begin{equation}
    \boxed{
    \sigma_{\rm CG}'
    =
    \frac{\Sigma'}{1-a\eta_{\rm CG}},
    \qquad
    p_s'
    =
    \frac{\eta_{\rm CG}\Sigma'}{1-a\eta_{\rm CG}}
    }.
\label{eq:app_cg_constitutive_derivatives}
\end{equation}
The point \(\eta_{\rm CG}=1/a\) is singular in this closure, so the regular
near-horizon solution continuously connected to the one below lies on
\(\eta_{\rm CG}<1/a\).

Differentiating Eq.~\eqref{eq:app_cg_force}, using
Eq.~\eqref{eq:app_cg_radial_pressure_derivative}, and substituting
Eq.~\eqref{eq:app_cg_constitutive_derivatives} gives
\begin{equation}
    \mathcal F_{{\rm CG},c}'
    =
    \frac{\Sigma_c'}{1-a\eta_{\rm CG}}
    \left(g_c+\eta_{\rm CG}h_c\right)
    +
    \sigma_{{\rm CG},c}g_c'
    +
    p_s h_c'
    .
\label{eq:app_cg_force_prime}
\end{equation}
This is Eq.~\eqref{eq:restoring_derivative}. To make the geometry explicit,
write again \(x=\sqrt\delta\). Then
\begin{equation}
    g_c=\frac{1-x^2}{4r_cx},
    \qquad
    h_c=\frac{1+x}{r_c},
\label{eq:app_cg_gh_values}
\end{equation}
while
\begin{equation}
    g_c'
    =
    -\frac{(1-x^2)(1+3x^2)}{8r_c^2x^3},
    \qquad
    h_c'
    =
    -\frac{(1+x)(3x-1)}{2r_c^2x}.
\label{eq:app_cg_gh_prime_values}
\end{equation}
Together with
\begin{equation}
    \Sigma_c'
    =
    \frac{\Sigma_c}{2r_cx^2}
    \left(1-9x^2\right),
\label{eq:app_cg_Sigmaprime_x}
\end{equation}
these equations contain all terms in the leading restoring derivative.
Setting Eq.~\eqref{eq:app_cg_force_prime} to zero and solving for the
constitutive response gives
\begin{equation}
    \boxed{
    \eta_{\rm CG}^{\rm crit}
    =
    -
    \frac{
        \Sigma_c'g_c
        +\sigma_{{\rm CG},c}g_c'
        +p_s h_c'
    }{
        \Sigma_c'h_c
        -a\left(
            \sigma_{{\rm CG},c}g_c'
            +p_s h_c'
        \right)
    }
    }.
\label{eq:app_cg_eta_exact}
\end{equation}
No Weyl-dominance approximation enters this result: the source relation follows
from the one-function consequence of the full Einstein--Weyl equations, while
the force balance uses the physical surface stress tensor.

For the near-horizon limit \(x=\sqrt\delta\to0^+\),
Eq.~\eqref{eq:app_cg_Sigma} gives
\begin{equation}
    \Sigma_c
    =
    \frac{8\alpha_g r_s}{r_c^4}x
    =
    O(x).
\label{eq:app_cg_Sigma_near_horizon}
\end{equation}
Expanding Eq.~\eqref{eq:app_cg_eta_exact} consistently in \(x\) gives
\begin{equation}
    \boxed{
    \eta_{\rm CG}^{\rm crit}
    =
    \frac{1}{a}
    -
    \frac{\Sigma_c}{a^2p_s}
    +
    O(\delta)
    =
    \frac{1}{a}
    -
    \frac{8\alpha_g r_s}{a^2p_s r_c^4}\sqrt\delta
    +
    O(\delta)
    }.
\label{eq:app_cg_eta_near_horizon}
\end{equation}
Evaluating the sign of Eq.~\eqref{eq:app_cg_force_prime} on the regular side
\(1-a\eta_{\rm CG}>0\) then gives the finite stable interval
\begin{equation}
    \boxed{
    \eta_{\rm CG}^{\rm crit}
    <
    \eta_{\rm CG}
    <
    \frac{1}{a}
    }.
\label{eq:app_cg_stable_interval}
\end{equation}
For integrated isotropy, \(a=1\), this reduces to
\begin{equation}
    1-
    \frac{8\alpha_g r_s}{p_s r_c^4}\sqrt\delta
    +O(\delta)
    <
    \eta_{\rm CG}
    <1,
    \qquad
    \eta_{\rm CG}^{\rm crit}\to1.
\label{eq:app_cg_isotropic_interval}
\end{equation}
Thus the constitutive response remains finite as the membrane approaches the
exterior Schwarzschild root, in direct contrast with
Eq.~\eqref{eq:app_israel_near_horizon}.

Finally, Eq.~\eqref{eq:app_cg_Sigmaprime} shows that
\begin{equation}
    \Sigma_c'=0
    \quad\Longleftrightarrow\quad
    \delta=\frac19
    \quad\Longleftrightarrow\quad
    r_c=\frac98r_s.
\label{eq:app_cg_endpoint}
\end{equation}
At this point \(x=1/3\), and
Eq.~\eqref{eq:app_cg_gh_prime_values} gives
\begin{equation}
    h_c'=0,
    \qquad
    g_c'=-\frac{4}{r_c^2}.
\end{equation}
The restoring derivative therefore loses all finite constitutive dependence:
\begin{equation}
    \mathcal F_{{\rm CG},c}'
    =
    -\frac{4\sigma_{{\rm CG},c}}{r_c^2}<0.
\label{eq:app_cg_endpoint_instability}
\end{equation}
No finite regular value of \(\eta_{\rm CG}\) stabilizes the leading radial
mode at the endpoint. Hence the finite-response near-horizon regime is confined
to the ultracompact side
\begin{equation}
    r_s<r_c<\frac98r_s,
\end{equation}
subject throughout to the constitutive stability condition. This location is
fully compatible with the physical interpretation of the solution: the
membrane is intended to represent a black-hole state and therefore belongs on
the ultracompact side of the familiar stellar Buchdahl radius. The common
value \(9r_s/8\) arises independently here from the zero of \(\Sigma_c'\),
which terminates the finite-response radial regime.

\bibliography{apssamp}

\end{document}